\documentclass[aps,prb,twocolumn]{revtex4}

\usepackage{graphicx}

\usepackage{latexsym}

\begin{document}

\setlength{\pdfpagewidth}{8.5in}
\setlength{\pdfpageheight}{11in}

\title{Hartree-Fock Computation of the High-T$_c$ Cuprate Phase Diagram}

\author{R. B. Laughlin}

\affiliation{Department of Physics, Stanford University,
Stanford, CA 94305}

\begin{abstract} 
A computation of the cuprate phase diagram is presented.  Adiabatic 
deformability back to the density function band structure plus symmetry 
constraints lead to a Fermi liquid theory with five interaction 
parameters.  Two of these are forced to zero by experiment.  The 
remaining three are fit to the moment of the antiferromagnetic state at 
half filling, the superconducting gap at optimal doping, and the maximum 
pseudogap. The latter is identified as orbital antiferromagnetism. 
Solution of the Hartree-Fock equations gives, in quantitative agreement 
with experiment, (1) quantum phase transitions at 5\% and 16\% $p$-type 
doping, (2) insulation below 5\%, (3) a $d$-wave pseudogap quasiparticle 
spectrum, (4) pseudogap and superconducting gap values as a function of 
doping, (5) superconducting $T_c$ versus doping, (6) London penetration 
depth versus doping, and (7) spin wave velocity.  The fit points to 
superexchange mediated by the bonding O atom in the Cu-O plane as the 
causative agent of all three ordering phenomena.
\end{abstract}

\homepage[R. B. Laughlin: ]{http://large.stanford.edu}

\date{\today}

\maketitle

\section{Introduction}

\begin{figure}
\includegraphics[scale=0.4]{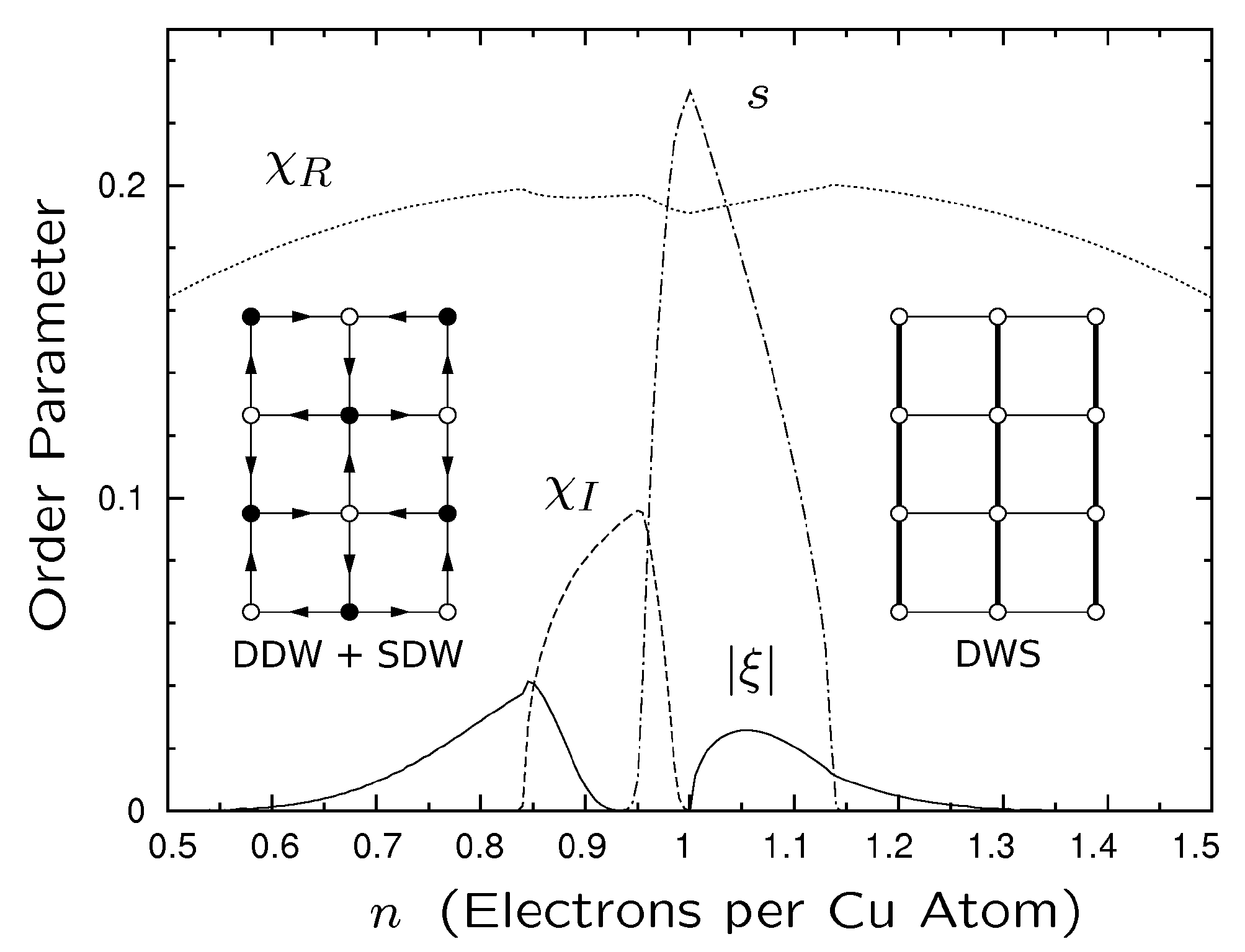}
\caption{Zero-temperature phase diagram of the Hamiltonian ${\cal H}_0 + 
\Delta {\cal H}$ defined by Eqs. (\ref{h0}) and (\ref{dh}) for the case 
of $U = 0.76 \, t$, $J = 0.75 \, t$, $V_c = 0.87 \, t$, $t' = 0.1 \, t$, 
and $V_n = V_t = 0$. The order parameters $s$, $\chi_I$, and $\xi$ are 
defined by Eqs. (\ref{opdef}). The self-consistency conditions are 
Eqs. (\ref{scn}) - (\ref{scxi}). The insets show the sense of the signs 
for spin and orbital current antiferromagnetism (left) and $d$-wave 
superconductivity (right). The system insulates everywhere $s$ is 
nonzero.}
\label{f1}
\end{figure}

The purpose of this paper is to discuss the theoretical phase diagram 
for the high-$T_c$ cuprate superconductors shown in Fig. \ref{f1} 
\cite{bonn,copper,broun,orenstein,mind,manifesto,wahl,rblprl}. It is 
generated using standard Hartree-Fock methods starting from a Fermi 
liquid theory with three interaction parameters \cite{agd,fetter,pines}. 
It is characterized by three interpenetrating order parameters: spin 
antiferromagnetism, or spin density wave (SDW), $d$-wave 
superconductivity (DWS) and orbital current antiferromagnetism, or 
$d$-density wave (DDW) \cite{ddw,nayak,dimov}.

However, the central issue of the paper is not building better models 
for the cuprates but the application of elementary quantum mechanics to 
them. The equations that generate Fig. \ref{f1} are not just made up. 
They are the {\it only} equations one can write down that are compatible 
with adiabatic evolution out of a fictitious metallic parent, plus a 
handful of experimental fiducials.  This evolution, which strictly 
enforces the Feynman rules, is the starting point of all conventional 
solid state physics. The ability of these equations to account broadly 
and well for all key aspects of high-$T_c$ phenomenology thus indicates 
that high-$T_c$ cuprates are {\it not} qualitatively different from 
other solids, as has often been suggested might be the case, but are 
simply materials with unusually complex low-energy spectroscopy. This 
complexity, which results from delicate interplay of multiple order 
parameters, has prevented the problem from being solved empirically.  
But elementary quantum mechanics so circumscribes the mathematics that 
one can say with confidence that the phase diagram in Fig. \ref{f1} is 
correct, even though one of its features, the identification of DDW with 
the cuprate pseudogap, is still in doubt phenomenologically 
\cite{ckn,stock,macridin,kee,macdougall}.

The results reported in this paper therefore have significance far 
greater than simply accounting for the behavior of a particular class of 
materials. The 25-year history of the cuprates has demonstrated rather 
brutally that first-principles theoretical control of solids at the 
energy scales relevant to electronic transport does not exist. This is 
so even though the underlying Hamiltonian of conventional matter is 
known exactly. Computers were not able to solve this problem. It was too 
hard for them. One obtains control, if at all, only by exploiting the 
universal low-energy properties of quantum phases.  The simple equations 
that describe these properties are the starting point for predictive 
computation. The practice of adiabatically evolving from fictional 
parent vacua is precisely what distinguishes solid state physics from 
materials chemistry, and is what makes it so much more powerful than the 
latter as a basis of engineering.

\subsection*{Historical Background}

The discovery of high-$T_c$ cuprate superconductivity revealed that 
standard methods for computing the properties of solids were more 
seriously flawed than previously thought 
\cite{bednorz,narrows,wu,maeda,schilling}. On the one hand, the 
materials in question were sufficiently conventional chemically that 
they should have yielded to ordinary self-consistent band structure 
analysis \cite{mattheiss}. The latter requires them to be metals in the 
absence of translational symmetry breaking. On the other, their 
phenomenology was totally incompatible with the conventional theory of 
metals \cite{metals}. Not only were their superconducting transition 
temperatures higher than existing theory said was possible without 
structural instability, their transport phenomenology was wildly 
irregular, and the violent variation of the superconducting transition 
temperature and superfluid density had no precedent 
\cite{cohen,marginal,takagi}.

Accordingly, Anderson and others suggested at the time that cuprate 
superconductivity might be an important new aspect of Mott insulation, 
an equally serious conceptual issue that had emerged 40 years earlier 
\cite{rvb,cyrot,mott}. This proposition was, and still is, extremely 
radical. Its central premise is that standard practices of solid state 
physics based on the adiabatic principle are irrelevant to these 
materials \cite{brainwash}. Nonetheless it has now become mainstream and 
central to the field, in part because so many experiments have defied 
conventional explanation. It has also given rise to a number of related 
non-adiabatic theoretical ideas such as the non-Fermi liquid state, the 
holographic metal and the orthogonal metal 
\cite{schofield,stewart,kirkpatrick,jiang,holographic,orthogonal}.

Unfortunately, the phenomenological definition of a Mott insulator has 
always been somewhat difficult to state and is sometimes expanded to 
include the entire class of ordinary magnetic insulators \cite{brandow}. 
The underlying idea is of a system that insulates when it ought not to. 
Thus the spin-unpolarized band structures of the transition metal 
monoxides FeO, CoO, and NiO are metallic but the materials themselves 
are all good insulators \cite{norman,ohta}. CoO has an odd number of 
electrons per unit cell.  All three oxides possess antiferromagnetic 
order at zero temperature, which doubles the unit cell and thus formally 
allows them to insulate by the usual rules of band structure 
\cite{slater,terakura, anisimov}. However, the strength of conventional 
exchange is inadequate to cause insulation in this way except in NiO, 
and all three materials continue to insulate above their Ne\`{e}l 
temperatures. Other materials typically (but not always) categorized as 
Mott insulators include MnO, V$_2$O$_3$, Fe$_2$O$_3$, LaTiO$_3$, 
Y$_2$Ru$_2$O$_7$, YTiO$_3$, YVO$_3$, and Sr$_2$VO$_4$ 
\cite{mcwhan,nakotte,pasternak,lee,park,patterson,ulrich,zhou}. The 
majority of identified Mott insulators are transition metal oxides.

The enormous amount of theoretical work stimulated by the cuprate 
discovery has now built up a strong case that the Mott insulator does 
not exist as a distinct zero-temperature state of matter.  This was 
arguably unclear when the cuprates were first discovered, but it is no 
more. Two decades of intense focus on the problem have not led to a 
single instance of an actual wavefunction for a Mott insulator written 
down in terms of the underlying electron coordinates 
\cite{imada,sachdev,pal,senthil,gebhard}. The resonating valence bond 
state of Anderson appears to be a counterexample, but this is not so 
\cite{vanilla,jain}. It is actually a $d$-wave superconductor. It is 
made by adding a short-range Coulomb repulsion to a superconducting 
Hamiltonian and then taking the strength of this repulsion to large 
values while legislating that no phase transition occurs 
\cite{gossamer}. Were such a perturbation actually applied without the 
unphysical constraint it would cause a phase transition to spin 
antiferromagnetism.  No numerical calculation based on a conventional 
Hamiltonian finds a resonating valence bond state 
\cite{dagotto,maier,edegger,leung}.

\section*{Experimental Constraints}

As materials and experimental techniques improved over time, the purely 
empirical case for a new quantum state incompatible with the theory of 
metals became progressively weaker. After several years of failure 
Josephson tunneling was finally observed between YBa$_2$C$_3$O$_{7 - 
\delta}$ and Pb, thus dispelling concerns that cuprate superconductivity 
might not be a traditional Cooper pair condensate \cite{sun,notoe}. 
Ideas about the non-fermionic nature of the superconductor's excitation 
spectrum were laid to rest by observation sharp fermionic quasiparticles 
in photoemission \cite{damascelli}. Controversy over the symmetry of the 
superconducting order parameter was settled in favor of $d$-wave pairing 
by observation of the node in photoemission, a sign change in Josephson 
tunneling, and half-integral trapped flux in magnetometery 
\cite{monthoux,littlewood,wollman,scalapino,tsuei}. Ideas about the 
non-existence of a Fermi surface were disproved by photoemission 
observation of Fermi surfaces in overdoped samples agreeing in detail 
with band structure and the Luttinger sum rule \cite{plate,luttinger}. 
And, finally, superconductors placed in magnetic fields strong enough to 
crush their superconductivity were found to exhibit quantum 
oscillations, thus demonstrating the presence of a Fermi surface at 
low-energy scales in the zero-temperature normal state 
\cite{hussey,sebastian,vignolle}.

The finite-temperature properties of the cuprates continued to be 
problematic, especially above the superconducting transition temperature 
near optimal doping \cite{batlogg}. However, as the temperatures were 
lowered to zero the behavior inevitably evolved into something simple 
and conventional, most notably when the superconductivity was suppressed 
with a magnetic field \cite{ando}. Repeated and consistent failure of 
the strange metal behavior to persist to low temperature has now 
demonstrated that it has nothing to do with quantum states of matter but 
is rather a critical phenomenon associated with a zero-temperature phase 
transition beneath the superconducting dome 
\cite{zaanen,kallin,mielke,butch}. The possibility that this phase 
transition is pseudogap development remains controversial 
\cite{varma,where}.

The occurrence of the pseudogap below optimal doping is associated with 
the reconstruction of the Fermi surface into pockets, as would be 
expected if a density wave had formed \cite{chakravarty}. The pseudogap, 
first discovered in magnetic resonance and optical conductivity, was 
later identified in photoemission as a $d$-wave quasiparticle dispersion 
that persisted above the superconducting transition temperature 
\cite{timusk,loeser}. Subsequent experiments revealed the existence of 
two $d$-wave gaps, one associated with the superconductivity and another 
antagonistic to it \cite{tacon,doiron,valenzuela,trisect}. Scanning 
tunneling microscopy has now shown that pseudogap has complex 
position-dependent structure that is inherently glassy 
\cite{rosch,kohsaka}. It has also shown that conventional fermionic 
quasiparticles exist in the presence of the pseudogap and that they have 
the ability to propagate coherently large distances through it and 
interfere \cite{mcelroy}.

\subsection*{Orbital Antiferromagnetism}

Several years before the quantum oscillation discovery, a group of us 
predicted that a reconstructed Fermi surface would appear when the 
superconductivity was destroyed by a strong magnetic field \cite{ddw}. 
We argued that conventional translational symmetry breaking had to be 
the cause of the pseudogap because nothing else could be written down as 
actual equations.  We proposed specifically that the pseudogap was the 
signature of orbital antiferromagnetism. Our grounds were (1) that there 
was no other way to account for a $d$-wave pseudogap that was compatible 
with the adiabatic principle and (2) that instability to such order was 
an unavoidable consequence of antiferromagnetic exchange stabilization 
of $d$-wave superconductivity out of a metallic parent. We named this 
order $d$-density wave (DDW) to distinguish it from the gauge theory 
flux vacua, which were mathematically similar but conceptually different 
\cite{affleck,liang}. However, after much searching the predicted 
magnetic Bragg peaks were not found, so the purely empirical case for 
the order could not be made \cite{dai}.

The subsequent discovery of quantum oscillations changed this situation.  
The original theoretical grounds for anticipating Fermi surface 
reconstruction had not changed, and attempts to reconcile it with the 
underlying quantum mechanics without doubling the unit cell proved 
impossible \cite{rice}. The only explanation compatible with the 
adiabatic principle is that DDW order is, in fact, present in the 
cuprates, and that failure to detect clean magnetic Bragg peaks from it 
has been a consequence of pseudogap glassiness.

There is nothing extraordinary about orbital magnetism from the point of 
view of quantum mechanics. The $^3P_2$ ground state of the neutral O 
atom has a total magnetic moment of $3 \mu_B$, $2 \mu_B$ of which comes 
from the spin and $1 \mu_B$ of which comes from the orbit.  Orbital 
antiferromagnetism is normally quenched in solids, but is it 
conceptually no different from spin antiferromagnetism \cite{schroeter}.

The failure to find signature magnetic Bragg peaks contributed 
materially to the development of the Varma current-loop theory, which 
has many similar features but does not break translation symmetry 
\cite{loop,loopmodes}. There is now some experimental support for this 
theory, although it is controversial 
\cite{fauque,mook,sonier,li,bourges,strassle}. However, the loop current 
insulating state has the same problem the Mott insulator does: It cannot 
be written down in electron coordinates.

Both kinds of spontaneous current would be difficult to detect in 
conventional Cu or O magnetic resonance in any cuprate because of 
symmetry \cite{lederer}.

\section{Adiabatic Evolution}

Let us now briefly review the idea of adiabatic evolution.  We imagine a 
fictitious Hamiltonian ${\cal H}_0$, usually noninteracting electrons 
moving in a periodic potential, that is easy to diagonalize. We then add 
a ``perturbation'' $\lambda ({\cal H} - {\cal H}_0)$, where ${\cal H}$ 
is the true Hamiltonian, and slowly increase $\lambda$ from 0 to 1.  
Each time we infinitesimally increment $\lambda$, one of two things 
happens: either (1) the ground state and low-lying excitations remain in 
one-to-one correspondence, or (2) they do not.  If the former is the 
case, we say that the system has remained in the same quantum phase.  If 
the latter is the case, we say that it has undergone a quantum phase 
transition \cite{shankar,sondhi}.

Adiabatic mapping is what enables universal characterization of 
low-energy excitations inside a given phase.  Thus, for example, when we 
say that electrons in a conventional metal behave as though they do not 
interact, we really mean that there is an adiabatic path back to ${\cal 
H}_0$ that does not encounter any phase transitions. Were there no such 
path, it would make no sense to talk about a metal's electrons and holes 
as abstractions, or to write down equations for them that involve only 
small effective interactions. The electrons in a conventional metal 
interact extremely strongly, as do electrons in solids generally. The 
correct Hamiltonian is always

\begin{displaymath}
{\cal H} = - \sum_j \frac{\hbar^2}{2m} \nabla_j^2 -
\sum_\alpha \frac{\hbar^2}{2M_\alpha} \nabla_\alpha^2 -
\sum_{j \alpha} \frac{Z_\alpha e^2}{|{\bf r}_j - {\bf R}_\alpha|}
\end{displaymath}

\begin{equation}
+ \sum_{j < k} \frac{e^2}{|{\bf r}_j - {\bf r}_k|}
+ \sum_{\alpha < \beta} \frac{Z_\alpha Z_\beta}
{|{\bf R}_\alpha - {\bf R}_\beta|}
\label{h}
\end{equation}

\noindent 
where ${\bf r}_j$ denotes the location of the $j$th electron and 
${\bf R}_\alpha$ denotes the location of an ion of mass $M_\alpha$. The 
kinetic and potential energies of the valence electrons are therefore 
always comparable by virtue of the virial theorem. It is not true 
that the Coulomb interactions in the cuprates are enormously bigger than 
they are in other solids, such as elemental Si or Na metal. They are
just the same.

Whether the interactions of Eq. (\ref{h}) are strong enough to 
destabilize the metallic state necessarily and always is an interesting 
question, but an academic one in light of the enormous body of 
experimental precedent in metals \cite{catastrophe,trash}. Moreover the 
idea that metals might be inherently unstable at low-energy scales does 
not in any way invalidate computational procedures based on adiabatic 
evolution from a fictitious noninteracting parent state, which is to 
say, sums of conventional metallic Feynman graphs \cite{schrieffer}.

It is not necessary that phase transitions should occur at isolated 
points in the interval $0 < \lambda < 1$, but this is usually the case.  
The reason is renormalization \cite{belitz}. As a system is made larger 
and larger its measured properties eventually begin to change in a 
deterministic way.  A renormalization fixed point is a Hamiltonian whose 
low-energy excitation spectrum stays the same when the system size is 
made larger.  If all the perturbations to this Hamiltonian diminish 
under renormalization, we say the fixed point is attractive, and we 
associate it with a stable state of matter. If at least one perturbation 
grows with renormalization, we say the fixed point is repulsive, and we 
associate it with a continuous phase transition. The difference between 
repulsion and attraction is why phases of matter occupy open sets of 
$\lambda$ values, while the transitions between them tend to occur at 
points.

The logical inconsistency of the Mott insulator concept is now easy to 
spot: It is perfectly reasonable that a system should pass through a new 
phase of matter on the way from $\lambda = 0$ (a metal) to $\lambda = 1$ 
(a $d$-wave superconductor) but one is obligated to say what it is.

\section{Fundamental Equations}

Correct equations for the cuprate problem are easy to write down once 
one accepts that they must describe adiabatic evolution out of the 
density functional band structure. The latter is described adequately by 
\cite{mattheiss}

\begin{displaymath}
{\cal H}_0 = - t \sum_{< \! j k \! >}^{2N} \sum_\sigma
( c_{j \sigma}^\dagger c_{k \sigma} +
c_{k \sigma}^\dagger c_{j \sigma} )
\end{displaymath}

\begin{equation}
+ t' \sum_{< \! j \ell \! >}^{2N} \sum_\sigma
( c_{j \sigma}^\dagger c_{\ell \sigma} +
c_{\ell \sigma}^\dagger c_{j \sigma} )
\label{h0}
\end{equation}

\noindent
where $< \! j k \! >$ denotes the set of near-neighbor pairs of $N$ 
sites on a planar square lattice of lattice constant $b$ and $< \! j 
\ell \! >$ denotes the set of second-neighbor pairs.  This description 
is inaccurate far from the fermi surface, but the high-energy 
excitations poorly described are not important.

\subsection*{Fermi Liquid Parameters} 

Each time one increments $\lambda$ the small perturbations renormalize 
to an effective Hamiltonian of the form

\begin{displaymath}
\Delta {\cal H}
= U \sum_j^N c_{j \uparrow}^\dagger c_{j \downarrow}^\dagger
c_{j \downarrow} c_{j \uparrow}
\end{displaymath}

\begin{displaymath}
+ \frac{J}{2} \sum_{< \! jk \! >}^{2N}
\sum_{\sigma \sigma'}  \biggl[
c_{j \sigma}^\dagger c_{k \sigma'}^\dagger
c_{k \sigma} c_{j \sigma'} 
- \frac{1}{2} c_{j \sigma}^\dagger c_{k \sigma'}^\dagger
c_{k \sigma'} c_{j \sigma} \biggr]
\end{displaymath}

\begin{displaymath}
+ V_t \sum_{< \! j k \! >}^{2N} \sum_{\sigma \sigma'}
(c_{j \sigma}^\dagger c_{k \sigma} + c_{ k \sigma}^\dagger c_{j \sigma})
(c_{j \sigma'}^\dagger c_{j \sigma'} + 
c_{k \sigma'}^\dagger c_{k \sigma'} - \frac{1}{2} )
\end{displaymath}

\begin{displaymath}
+ V_n \, \sum_{<jk>}^{2N} \sum_{\sigma \sigma'}
\, c_{j \sigma}^\dagger c_{k \sigma '}^\dagger
c_{k \sigma '} c_{j \sigma} 
\end{displaymath}

\begin{equation}
+ V_c \, \sum_{<jk>}^{2N} \biggl[
c_{j \uparrow}^\dagger c_{j \downarrow}^\dagger
c_{k \downarrow} c_{k \uparrow} +
c_{k \uparrow}^\dagger c_{k \downarrow}^\dagger
c_{j \downarrow} c_{j \uparrow} \biggr] 
\label{dh}
\end{equation}

\noindent 
This represents the most general set of fermi liquid parameters allowed 
by symmetry.  Only pairwise interactions are relevant because the 
perturbation excites a quantum-mechanical gas of quasiparticles that is 
dilute.  Only lattice terms closer than second neighbors are relevant 
because these exhaust the low angular momentum scattering channels.  All 
terms associated with bonds must be rotationally invariant about the 
bond axis and reflection symmetric.  All terms must be spin-rotationally 
invariant and time-reversal syymmetric.  The parameters can, in 
principle, be energy-dependent, as they are, for example, if they are 
mediated by phonons, but this is irrelevant at the lowest energy scales.  
The difference between phonon-mediated pairing and purely electronic 
pairing is retardation, and this shows up only in high-energy 
spectroscopy \cite{schrieffer}. The parameters can also be 
doping dependent, but I find that they are not.

All the various parts of ${\cal H}$ added by incrementing $\lambda$ 
renormalize into fermi liquid parameters by definition if the system has 
not yet made a phase transition out of the metallic state 
\cite{abrikosov,nozieres,wolfle,depuis}. But they also do if a phase 
transition has occurred along the way that is mild. This is why 
conventional superconductors may be described simply with Feynman graph 
sums \cite{schrieffer}. The new state is still a quantum-mechanical 
combination of electron and holes of the parent metal.

Since ${\cal H} + \Delta {\cal H}$ is the most general effective 
Hamiltonian possible, {\it all} of the behaviors of the cuprate 
superconductors must be contained in it. There is no other possibility. 

\subsection*{Band Rigidity}

We may immediately set $V_t$ to zero.  Were it present its main effect 
would be to renormalize the kinetic energy in the doping-dependent way

\begin{equation}
t \rightarrow t + V_t \, \sum_{\sigma} \biggl[
< \! c_{j \sigma}^\dagger c_{j \sigma} \! >
+ < \! c_{k \sigma}^\dagger c_{k \sigma} \! > - \frac{1}{2}  \biggr]
\end{equation}

\noindent
where $< >$ denotes ground state expectation value. However, 
angle-resolved photoemission measurements find the asymptotic nodal 
fermi velocity to be $\hbar v_F = 2.0$ eV \AA $\,$ for both $p$-type and 
$n$-type materials, regardless of doping \cite{vishik,johnson,armitage}.

The absence of such a term makes sense physically. Doping dependence of 
$t$ simply means that the potential barrier through which the electrons 
tunnel depends on electron density.  Such dependence is already taken 
care of in the band structure through self-consistency.

The observed Fermi velocity requires $t = 0.19$ eV, which is about half 
the value predicted by the native band structure \cite{mattheiss}. The 
value of $t' = 0.1 \, t$ is also fixed by photoemission, although it is 
also consistent with calculations \cite{pavarini}. 

\section{Hartree-Fock Solution}

\label{hartreefock}

The ground state $| \Psi \! >$ is characterized by the 
expectation values

\begin{displaymath}
< \! c_{j \sigma}^\dagger c_{k \sigma} \! > =
\chi_R \pm i \, \chi_I 
\; \; \; \; \; \; \;
< \! c_{j \sigma}^\dagger c_{\ell \sigma} \! > =
\chi_R'
\end{displaymath}

\begin{equation}
< \! c_{j \uparrow}^\dagger c_{k \downarrow}^\dagger \! > 
=  \xi
\; \; \; \; \; \; \;
< \! c_{j \sigma}^\dagger c_{j \sigma} \! > 
= \frac{n}{2} \pm \, (-1)^\sigma s
\label{opdef}
\end{equation}

\noindent
with the signs as in Fig. \ref{f1}.  The order parameter $\xi$ describes 
$d$-wave superconductivity (DWS). The order parameter $\chi_I$ describes 
orbital antiferromagnetism (DDW). The order parameter $s$ describes spin 
antiferromagnetism (SDW). $\chi_R$ and $\chi_R'$ are not order 
parameters but measures of the ground state kinetic energy. $n$ is the 
site occupancy.

To simplify the calculation we constrain both $\chi_I$ and $s$ to 
be periodic in the doubled unit cell. This creates a mild artifact of 
allowing the system to conduct electricity at all nonzero dopings.  If 
this constraint is relaxed, SDW domain walls form, trapping carriers and 
causing the system to insulate everywhere SDW order is developed 
\cite{schulz, inui,sarkar,chubukov,gong}. The similar problem with DDW, 
while important, is less severe because (1) the DDW quasiparticle 
spectrum is gapless at half-filling and (2) the DDW symmetry breaking is 
discrete.

The Hartree-Fock approximation is expressed either as a sum of rainbow 
Feynman graphs or as a single Slater determinant variational ground 
state \cite{fetter}. In either case, each electron effectively moves in 
an average field generated by all the others. The variational ground 
state $| \Psi \! >$ takes the form

\begin{equation}
| \Psi \! > = \prod_{{\bf q} \nu} ( u_{{\bf q} \nu}
+ v_{{\bf q} \nu} \, c_{{\bf q} \nu \uparrow}^\dagger
c_{-{\bf q} \nu \downarrow}^\dagger ) | 0 \! >
\label{bcs}
\end{equation}

\noindent
where

\begin{equation}
c_{{\bf q} \nu \sigma}^\dagger 
= \sqrt{\frac{2}{N}} \, \sum_j^N
(a_{j \sigma}^{{\bf q} \nu})^* \, 
\exp(i {\bf q} \cdot {\bf r}_j ) \, c_{j \sigma}^\dagger
\label{c}
\end{equation}

\noindent 
is the creation operator for an electron of crystal momentum ${\bf q}$ 
and spin $\sigma$ in band $\nu$. The index $\nu$ is required because 
both kinds of antiferromagnetism double the unit cell. As usual, the 
vector ${\bf r}_j$ denotes the location of the $j$th site. The 
coefficients $a_{j \sigma}^{{\bf q} \nu}$ are the same in every unit 
cell and satisfy

\begin{equation}
\begingroup
\renewcommand*{\arraystretch}{1.5}
\left[ \begin{array}{cc}
\tilde{\Delta} & \tilde{\epsilon}_{\bf q} + i \tilde{\Delta}_{\bf q} \\
\tilde{\epsilon}_{\bf q} - i \tilde{\Delta}_{\bf q} & - \tilde{\Delta}
\end{array} \right]
\left[ \begin{array}{c}
a_{1 \uparrow}^{{\bf q} \pm} \\
a_{2 \uparrow}^{{\bf q} \pm}
\end{array} \right]
= \pm \tilde{E}_{\bf q} \, 
\left[ \begin{array}{c}
a_{1 \uparrow}^{{\bf q} \pm} \\
a_{2 \uparrow}^{{\bf q} \pm}
\end{array} \right]
\endgroup
\label{bandup}
\end{equation}

\begin{equation}
\begingroup
\renewcommand*{\arraystretch}{1.5}
\left[ \begin{array}{cc}
-\tilde{\Delta} & \tilde{\epsilon}_{\bf q} + i \tilde{\Delta}_{\bf q} \\
\tilde{\epsilon}_{\bf q} - i \tilde{\Delta}_{\bf q} & \tilde{\Delta}\\
\end{array} \right]
\left[ \begin{array}{c}
a_{1 \downarrow}^{{\bf q} \pm} \\
a_{2 \downarrow}^{{\bf q} \pm}
\end{array} \right]
= \pm \tilde{E}_{\bf q} \, 
\left[ \begin{array}{c}
a_{1 \downarrow}^{{\bf q} \pm} \\
a_{2 \downarrow}^{{\bf q} \pm}
\end{array} \right]
\endgroup
\label{banddown}
\end{equation}

\noindent
where

\begin{equation}
\tilde{\epsilon}_{\bf q} = - \, \biggl[ t + (\frac{3}{4} J + V_n - V_c) 
\chi_R \biggr] \biggl[ \cos(q_x) + \cos(q_y) \biggr]
\end{equation}

\begin{equation}
\tilde{t} =  t + (\frac{3}{4} J + V_n - V_c) 
\chi_R 
\label{tildet}
\end{equation}

\begin{equation}
\tilde{\Delta}_{\bf q} = ( \frac{3}{4} J + V_n + V_c) \chi_I
\biggl[ \cos(q_x) - \cos(q_y) \biggr]
\end{equation}

\begin{equation}
\tilde{\Delta} = ( U + 2 J) \, s
\end{equation}

\begin{equation}
\tilde{E}_{\bf q} = \sqrt{\tilde{\epsilon}_{\bf q}^2 + \tilde{\Delta}^2
+ \tilde{\Delta}_{\bf q}^2}
\end{equation}

\noindent
in units for which the bond length $b$ equals 1.

The absence of band mixing in Eq. (\ref{bcs}) is a consequence of the 
system's special symmetries. Combining Eq. (\ref{bandup}) with the 
complex conjugate of Eq. (\ref{banddown}) in the presence of $d$-wave
superconductivity, we obtain the Nambu matrix

\begin{equation}
{\cal H}_{\bf q} =
\left[ \begin{array}{cccc}
\tilde{\Delta} - \tilde{\mu} & 
\tilde{\epsilon}_{\bf q} + i \tilde{\Delta}_{\bf q} &
0 & \tilde{\theta}_{\bf q}^* \\
\tilde{\epsilon}_{\bf q} - i \tilde{\Delta}_{\bf q} &
- \tilde{\Delta} - \tilde{\mu} &
\tilde{\theta}_{\bf q}^* & 0\\
0 & \tilde{\theta}_{\bf q} &
\tilde{\mu} + \tilde{\Delta} & 
\! \! \! \!
- \tilde{\epsilon}_{\bf q} + i \tilde{\Delta}_{\bf q} \\
\tilde{\theta}_{\bf q} & 0 &
\! \! \! \! \! \!
- \tilde{\epsilon}_{\bf q} - i \tilde{\Delta}_{\bf q} & 
\tilde{\mu} - \tilde{\Delta} \\
\end{array} \right]
\end{equation}

\noindent
where

\begin{equation}
\tilde{\theta}_{\bf q} = 
(\frac{3}{4} J - V_n) \, \xi \, \biggl[
\cos(q_x) - \cos(q_y) \biggr]
\end{equation}

\begin{equation}
\tilde{\mu}_{\bf q} = \mu - 2 t' \biggl[
\cos(q_x + q_y) + \cos (q_x - q_y) \biggr]
\end{equation}

\noindent
with $\mu$ as the chemical potential. However since

\begin{displaymath}
\left[ \begin{array}{cccc}
1 & 0 & 0 & 0 \\
0 & 1 & 0 & 0 \\
0 & 0 & 0 & 1 \\
0 & 0 & 1 & 0 \\
\end{array} \right]
\; {\cal H}_{\bf q} \;
\left[ \begin{array}{cccc}
1 & 0 & 0 & 0 \\
0 & 1 & 0 & 0 \\
0 & 0 & 0 & 1 \\
0 & 0 & 1 & 0 \\
\end{array} \right]
\end{displaymath}

\begin{equation}
=
\left[ \begin{array}{cccc}
\tilde{\Delta} - \tilde{\mu}_{\bf q} & 
\tilde{\epsilon}_{\bf q} + i \tilde{\Delta}_{\bf q} &
\tilde{\theta}_{\bf q}^* & 0\\
\tilde{\epsilon}_{\bf q} - i \tilde{\Delta}_{\bf q} &
- \tilde{\Delta} - \tilde{\mu}_{\bf q} &
0 & \tilde{\theta}_{\bf q}^* \\
\tilde{\theta}_{\bf q} & 0 &
\tilde{\mu}_{\bf q} - \tilde{\Delta} & 
\! \! 
- \tilde{\epsilon}_{\bf q} - i \tilde{\Delta}_{\bf q} \\
0 & \tilde{\theta}_{\bf q} &
\! \! \! \! \! \!
- \tilde{\epsilon}_{\bf q} + i \tilde{\Delta}_{\bf q} &
\tilde{\mu}{\bf q} + \tilde{\Delta} \\
\end{array} \right]
\end{equation}

\noindent
the eigenvalues of ${\cal H}_{\bf q}$ are $\pm E_{\bf q}^\pm$, where

\begin{equation}
E_{\bf q}^\pm = \sqrt{(\pm \tilde{E}_{\bf q} - \tilde{\mu}_{\bf q})^2
+ |\tilde{\theta}_{\bf q}|^2} 
\end{equation}

\noindent
The bands therefore do not mix, and we have

\begin{equation}
u_{{\bf q} \pm} = \sqrt{\frac{E_{\bf q}^\pm + 
(\pm \tilde{E}_{\bf q} - \tilde{\mu}_{\bf q})}
{2 E_{\bf q}^\pm}}
\end{equation}

\begin{equation}
v_{{\bf q} \pm} = \sqrt{\frac{E_{\bf q}^\pm 
- (\pm \tilde{E}_{\bf q} - \tilde{\mu}_{\bf q})}
{2 E_{\bf q}^\pm}}
\end{equation}

\subsection*{Quasiparticle Energies}

The operators

\begin{equation}
b_{{\bf q} \nu \uparrow} 
= u_{{\bf q} \nu} \, c_{{\bf q} \nu \uparrow}
+ v_{{\bf q} \nu} \, c_{{\bf q} \nu \downarrow}^\dagger
\end{equation}

\begin{equation}
b_{{\bf q} \nu \downarrow} 
= u_{{\bf q} \nu} \, c_{{\bf q} \nu \downarrow}
- v_{{\bf q} \nu} \, c_{{\bf q} \nu \uparrow}^\dagger
\end{equation}

\noindent
destroy $| \Psi \! >$.  Their adjoints create quasiparticles of energy
$E_{\bf q}^\nu$.

\subsection*{Pairwise Contractions}

With a variational ground state of the form of Eq. (\ref{bcs}),
the expected interaction energy is the sum of all pairwise contractions.
Thus, for example, the second term of Eq. (\ref{dh}) gives a ground
state expectation value of

\begin{displaymath}
\sum_{\sigma \sigma'}  
< \! \Psi \biggl|
c_{j \sigma}^\dagger c_{k \sigma'}^\dagger
c_{k \sigma} c_{j \sigma'} 
- \frac{1}{2} c_{j \sigma}^\dagger c_{k \sigma'}^\dagger
c_{k \sigma'} c_{j \sigma} 
\biggr| \Psi \! >
\end{displaymath}

\begin{displaymath}
= - \frac{3}{2} \biggl\{
< \! c_{j \uparrow}^\dagger c_{k \downarrow}^\dagger \! >
\, < \! c_{k \downarrow} c_{j \uparrow} \! >
+ < \! c_{j \downarrow}^\dagger c_{k \uparrow}^\dagger \! >
\, < \! c_{k \uparrow} c_{j \downarrow} \! > \biggr\}
\end{displaymath}

\begin{displaymath}
- \sum_{\sigma \sigma'} (1 - \frac{1}{2} \delta_{\sigma \sigma'}) \;
< \! c_{j \sigma}^\dagger c_{k \sigma} \! > \,
< \! c_{k \sigma'}^\dagger c_{j \sigma'} \! >
\end{displaymath}

\begin{displaymath}
+ \sum_{\sigma \sigma'} (\delta_{\sigma \sigma'} - \frac{1}{2}) \;
< \! c_{j \sigma}^\dagger c_{j \sigma} \! > \,
< \! c_{k \sigma'}^\dagger c_{k \sigma'} \! >
\end{displaymath}

\begin{equation}
= - 3 | \xi |^2 - 3 (\chi_R + i \chi_I ) (\chi_R - i \chi_I )
- 2 s^2
\label{jterm}
\end{equation}

\noindent
Thus this term stabilizes all three kinds of order. Here we have used the
relations

\begin{equation}
< \! c_{j \uparrow}^\dagger c_{k \downarrow}^\dagger \! > =
< \! c_{k \uparrow}^\dagger c_{j \downarrow}^\dagger \! > =
< \! c_{k \downarrow} c_{j \uparrow} \! >^* =
< \! c_{j \downarrow} c_{k \uparrow} \! >^*
\end{equation}

\begin{equation}
< \! c_{j \uparrow}^\dagger c_{k \uparrow} \! > =
< \! c_{j \downarrow}^\dagger c_{k \downarrow} \! > =
< \! c_{k \uparrow}^\dagger c_{j \uparrow} \! >^* =
< \! c_{k \downarrow}^\dagger c_{j \downarrow} \! >^*
\end{equation}

\noindent
implicit in Eq. (\ref{bcs}).  

The last term in Eq. (\ref{dh}) gives

\begin{displaymath}
< \! \Psi | \,
c_{j \uparrow}^\dagger c_{j \downarrow}^\dagger
c_{k \downarrow} c_{k \uparrow} +
c_{k \uparrow}^\dagger c_{k \downarrow}^\dagger
c_{j \downarrow} c_{j \uparrow} \, | \Psi \! >
\end{displaymath}

\begin{displaymath}
= < \! c_{j \uparrow}^\dagger c_{k \uparrow} \! > \,
< \! c_{j \downarrow}^\dagger c_{k \downarrow} \! >
+ < \! c_{k \uparrow}^\dagger c_{j \uparrow} \! > \,
< \! c_{k \downarrow}^\dagger c_{j \downarrow} \! >
\end{displaymath}

\begin{equation}
= (\chi_R + i \chi_I)^2 + (\chi_R - i \chi_I)^2
\end{equation}

\noindent
It thus has no effect on either DWS or SDW but stabilizes DDW. 

The remaining two ``coulombic'' terms in Eq. (\ref{dh}) give

\begin{displaymath}
\sum_{\sigma \sigma'} < \! \Psi | \,
c_{j \sigma}^\dagger c_{k \sigma'}^\dagger 
c_{k \sigma'} c_{j \sigma} | \Psi \! >
\end{displaymath}

\begin{equation}
= 2 | \xi |^2 - 2 (\chi_R + i \chi_I) (\chi_R - i \chi_I)
+ n^2
\end{equation}

\noindent
per Eq. (\ref{jterm}) and

\begin{displaymath}
< \! \Psi | c_{j \uparrow}^\dagger c_{j \downarrow}^\dagger
c_{j \downarrow} c_{j \uparrow} | \Psi \! >
= < \! c_{j \uparrow}^\dagger c_{j \uparrow} \! >
\, < \! c_{j \downarrow}^\dagger c_{j \downarrow} \! >
\end{displaymath} 

\begin{equation}
= (\frac{n}{2})^2 - s^2
\end{equation}

\subsection*{Variational Energy}

Thus the total expected ground state energy is

\begin{displaymath}
\frac{1}{N} \,
\frac{< \! \Psi | \, {\cal H}_0 + \Delta{\cal H} \, | \Psi \! >}
{< \! \Psi | \Psi \! >} 
= - 8 t \chi_R + 8 t' \chi_R'
\end{displaymath}

\begin{displaymath} 
+ \Bigl[ \frac{n^2}{4} - s^2 \Bigr] \; U 
+ \Bigl[ 2 n^2 + 4 | \xi |^2 - 4 \chi_R^2 - 4 \chi_I^2 \Bigr] \, V_n
\end{displaymath}

\begin{equation}
- \Bigl[ 3 \chi_R^2 + 3 \chi_I^2 + 3 |\xi|^2 
+ 2 \, s^2 \Bigr] \, J + 
\Bigl[ 4 \chi_R^2 - 4 \chi_I^2 \Bigr] \, V_c
\end{equation}

\subsection*{Self-Consistency Equations}

The extremal condition

\begin{equation}
\delta \, \biggl\{
\frac{< \! \Psi | {\cal H}_0 + \Delta {\cal H} | \Psi \! >}
{< \! \Psi | \Psi \! >} \biggr\} = 0
\end{equation}

\noindent
then gives the equations

\begin{equation}
n = \frac{1}{8 \pi^2} \sum_\pm
\int_{-\pi}^\pi \int_{-\pi}^\pi \biggl[ 1 -
\frac{(\pm \tilde{E}_{\bf q} - \tilde{\mu}_{\bf q})}{E_{\bf q}^\pm} \biggr]
\, dq_x dq_y
\label{scn}
\end{equation}

\begin{displaymath}
s = - \frac{1}{(U + 2 J) s} \times \frac{1}{16 \pi^2} \sum_\pm
\end{displaymath}

\begin{equation}
\times \int_{-\pi}^\pi \int_{-\pi}^\pi 
\frac{\tilde{\Delta}^2}{\pm \tilde{E_{\bf q}}}
\biggl[ 1 -
\frac{(\pm \tilde{E}_{\bf q} - \tilde{\mu}_{\bf q})}{E_{\bf q}^\pm}
\biggr]
\, dq_x dq_y
\label{scs}
\end{equation}

\begin{displaymath}
\chi_R = - \frac{1}{\tilde{t}} 
\times 
(\frac{1}{64 \pi^2}) \sum_\pm
\end{displaymath}
 
\begin{equation}
\times 
\int_{-\pi}^\pi \int_{-\pi}^\pi
\frac{\tilde{\epsilon}_{\bf q}^2}{\pm\tilde{E}_{\bf q}} \biggl[
1 - \frac{( \pm \tilde{E}_{\bf q} - \tilde{\mu}_{\bf q})}
{E_{\bf q}^\pm} \biggr] \;
dq_x dq_y
\label{sccr}
\end{equation}

\begin{displaymath}
\chi_I = - \, \frac{1}{(3 J/4 + V_n + V_c ) \chi_I } 
\times (\frac{1}{64 \pi^2}) \sum_\pm
\end{displaymath}

\begin{equation}
\times \int_{-\pi}^\pi \int_{-\pi}^\pi
\frac{\tilde{\Delta}_{\bf q}^2}{\pm\tilde{E}_{\bf q}} \biggl[
1 - \frac{(\pm \tilde{E}_{\bf q} - \tilde{\mu}_{\bf q})}
{E_{\bf q}^\pm} \biggr] \;
dq_x dq_y
\label{scci}
\end{equation}

\begin{displaymath}
\xi = \frac{1}{ (3 J/4 - V_n ) \xi } \times
(\frac{1}{64 \pi^2}) \sum_\pm
\end{displaymath}

\begin{equation}
\times \int_{-\pi}^\pi \int_{-\pi}^\pi 
\frac{\tilde{\theta}_{\bf q}^2}{E_{\bf q}^\pm}
\; dq_x dq_y
\label{scxi}
\end{equation}

\noindent
Iterating these to convergence locates the minimum variational ground 
state energy

\section{Limiting Cases}

As a preliminary to fitting the parameters $U$, $J$, $V_c$ and $V_n$ to 
experiment we shall consider a handful of limiting cases.

\begin{figure}
\includegraphics[scale=0.4]{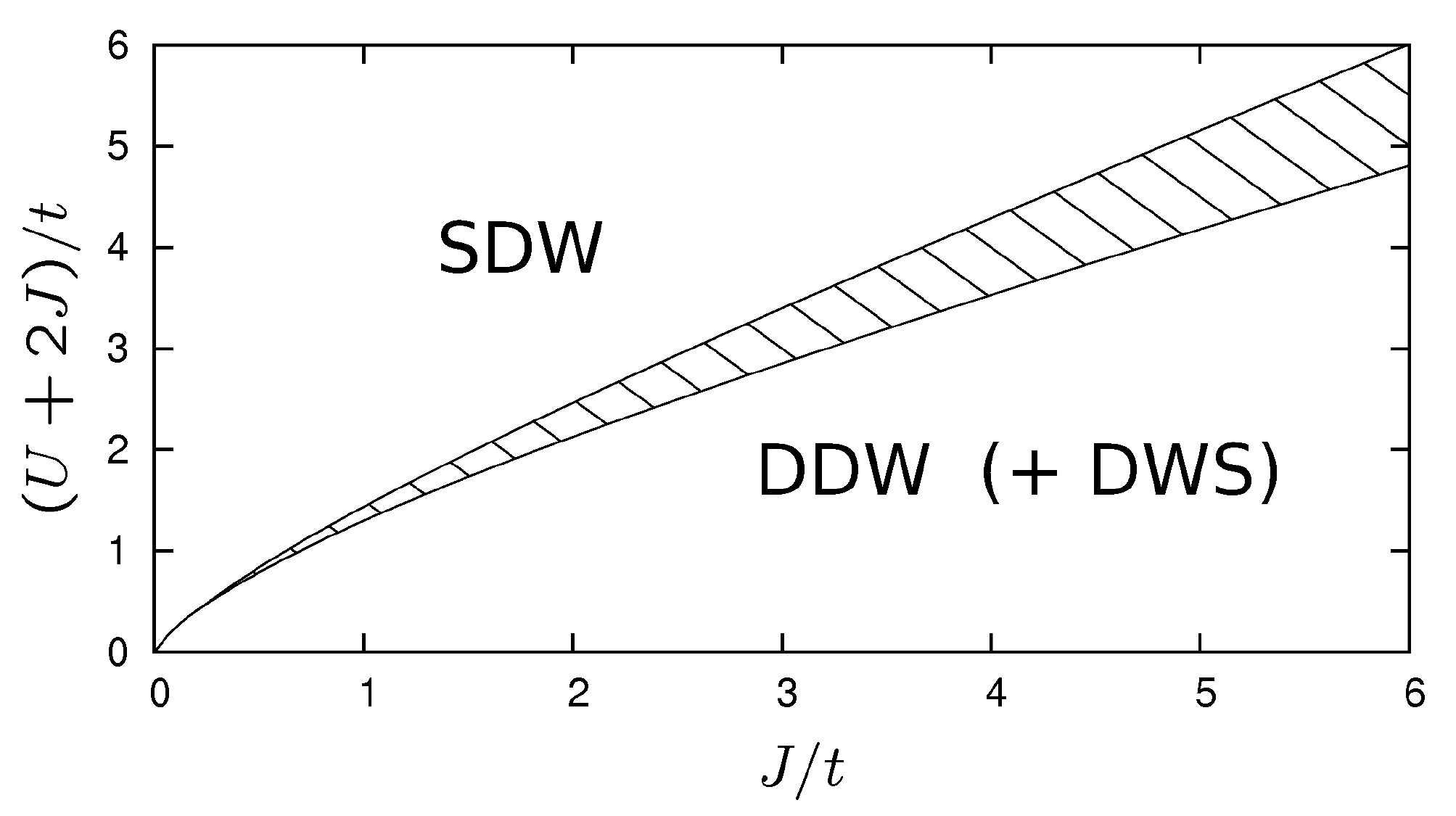}
\caption{Phase diagram for the particle-hole symmetric ($t' = 0$) case 
at half filling ($n = 1$) given by Eqs. (\ref{scn}) - (\ref{scxi}) with 
all parameters except $t$, $U$, and $J$ set to zero and with 
superconductivity suppressed by artificially holding $\xi$ to zero. The 
quantum phase transition between SDW and DDW is second order with a 
small region of coexistence (hashed). DDW is exactly degenerate with DWS 
at half-filling when only $U$ and $J$ are nonzero.  This is a 
consequence of a special symmetry described in Appendix \ref{phtrans}.}
\label{f2}
\end{figure}

\subsection*{Half-Filling Antiferromagnetism}

Let us first set all the parameters except $t$, $J$, and $U$ to zero and 
adjust the chemical potential $\mu$ to half-filling ($n=1$). The 
superconductivity is problematic in this limit, so let us also force 
$\xi = 0$ by hand. Iterating Eqs. (\ref{scn}) - (\ref{scxi}) to 
self-consistency, we obtain the result shown in Fig. \ref{f2}. There 
are two second-order phase transitions bounding a region of coexistence 
between spin antiferromagnetism (spin density wave SDW) and orbital 
antiferromagnetism ($d$-density wave DDW).

It is immediately clear from this figure that pure exchange 
characterized by $J$ tends to stabilize SDW and DDW equally. For $J > 
0.2 \, t$ one needs $U < 0$, which is unphysical, to achieve 
coexistence.  But the more important observation is that the requisite 
$U$ is relatively small.  The boundary in question is also actually 
multicritical. DDW and DWS are exactly degenerate in this limit. This 
may be seen by comparing Eqs. (\ref{scci}) and (\ref{scxi}), but the 
actual cause is a special symmetry described in Appendix \ref{phtrans}.

This result thus shows that strongly correlated superexchange ($J = 4 
t^2/U << U$) cannot account quantitatively for DWS in the cuprates 
\cite{superex}. One of the key experimental features of these materials 
is that small amounts of doping, a delicate perturbation, can violently 
disrupt the SDW and completely replace it with DWS. This implies that 
the system lies near a phase boundary. But Fig. \ref{f2} shows that the 
purely magnetic system cannot be near the phase boundary between SDW and 
DWS unless $U$ is small. This observation is backed up by the large body 
of numerical work on the Hubbard model, which shows that it does not 
account well for properties of the cuprates \cite{hubbard}.

\begin{figure}
\includegraphics[scale=0.4]{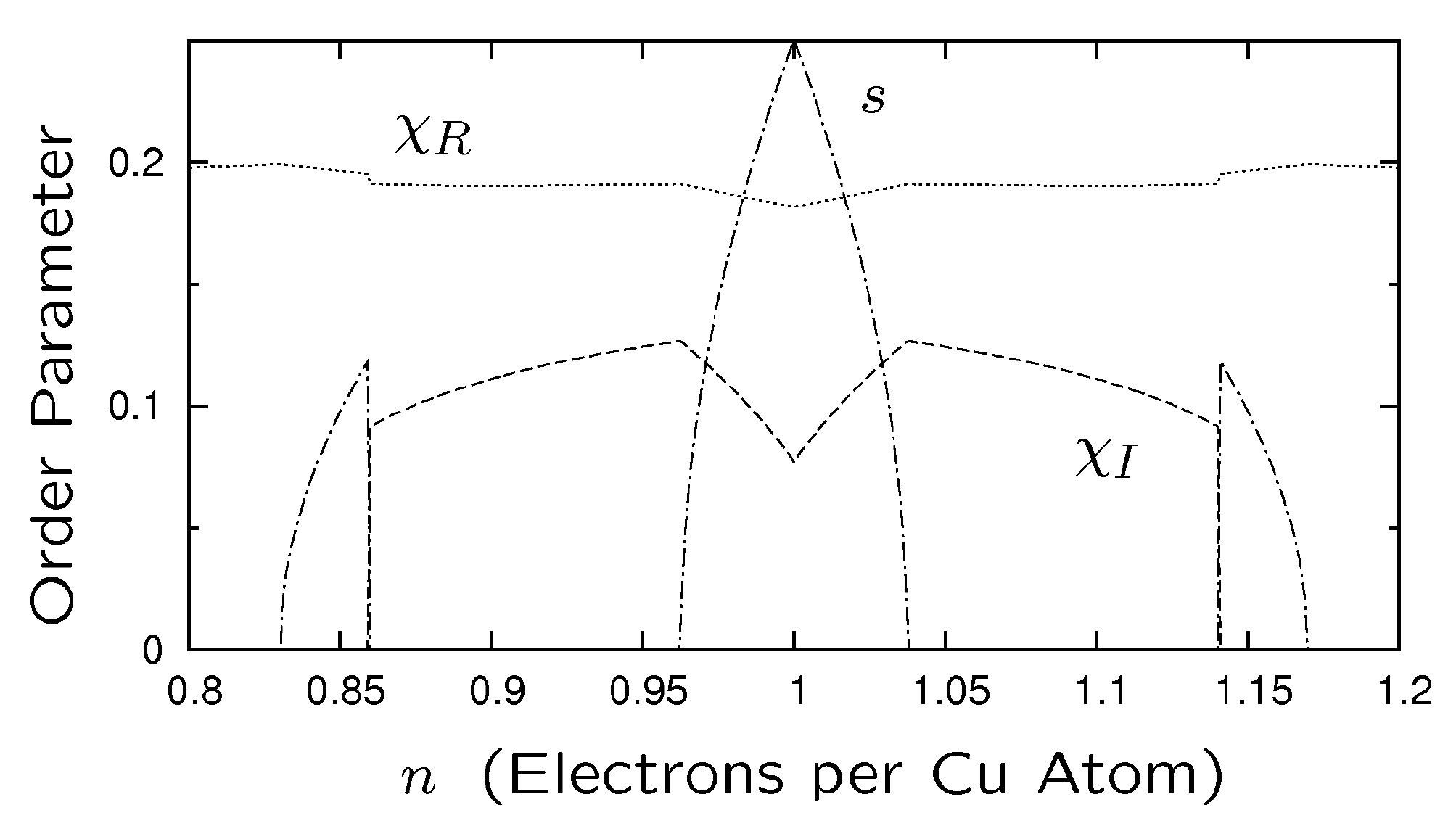}
\caption{Order parameters $s$ and $\chi_I$ defined by Eqs. 
(\ref{opdef}) as a function of $n$ obtained by solving Eqs. (\ref{scn}) 
- (\ref{scxi}) with $\xi$ artificially held to zero for the case of $U + 
2 J = 5.7 \, t$, $J = 5.9 \, t$, and $t' = V_c = V_n = 0$.  The kinetic 
energy parameter $\chi_R$ is also plotted. The parameters are 
determined by the conditions (1) $s = 0.25$ at $n = 1$ and (2) the 
system lie near the upper edge of the coexistence curve of Fig. 
\ref{f2}.  The occurrence of phase transitions at 4\% and 14\% and 
16\% doping is a general and robust consequence of these two 
constraints.  The re-entrance at 14\% is first order.} 
\label{f3} 
\end{figure}

\subsection*{SDW vs. DDW}

\begin{figure}
\includegraphics[scale=0.55]{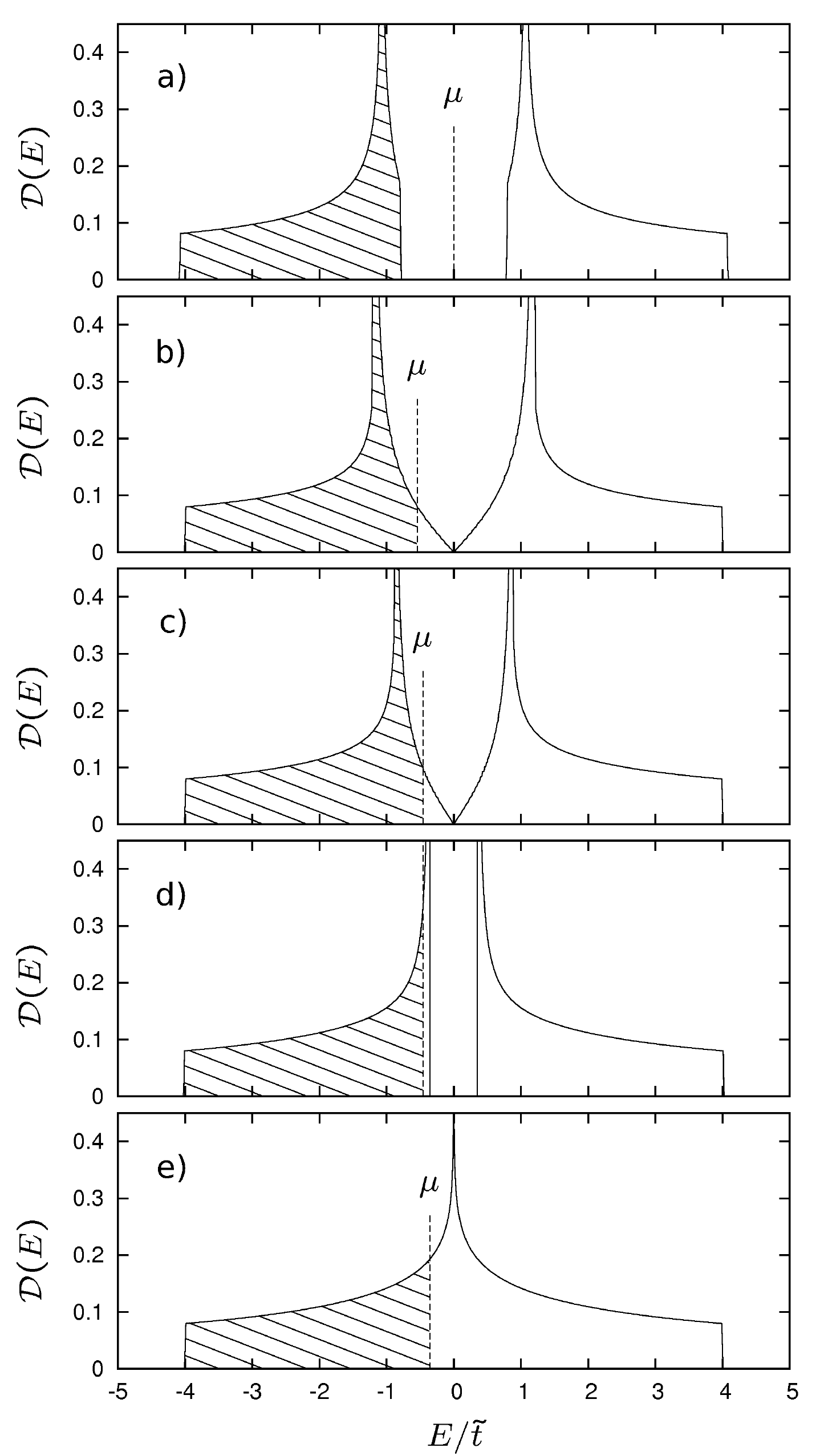}
\caption{Density of states ${\cal D}(E)$ defined by Eq. (\ref{dos}) 
for the calculation of Fig. \ref{f3} at the doping densities (a) $n = 
1.000$, (b) $n = 0.962$, (c) $n = 0.860$, (d) $n = 0.859$, and (e) $n = 
0.831$. The energy unit is $\tilde{t}$ defined by Eq. (\ref{tildet}).  The
shaded region is occupied.  The chemical potential $\mu$ is the same in
(c) and (d), as required at a first-order transition.}
\label{f4} 
\end{figure}

Let us next relax the condition that $n = 1$ and repeat the calculation 
of Fig. \ref{f2}, again artificially forcing $\xi = 0$, with 
fixed values of $U + 2 J = 5.7 \, t$ and $J = 5.9 \, t$.  These 
parameters are unphysically large, in part because $\tilde{t}/t \simeq 1.8$,
per Eq. (\ref{tildet}). They are fit to the conditions that 
(1) the system lie near the upper edge of the coexistence region in Fig. 
\ref{f2} and (2) the spin moment at half-filling be $s=0.25$, half the 
maximum classically allowed value. A moment of this size is 
characteristic of the insulating cuprates 
\cite{yamada,vaknin,regnault,tranquada}. The result, shown in Fig. 
\ref{f3}, reveals that the SDW is destroyed by a doping of 4\%, a number 
characteristic of spin antiferromagnetism disappearance of the cuprates. 
SDW is supplanted at this density by DDW, a pseudogap candidate with a 
$d$-wave quasiparticle spectrum. DDW itself then becomes unstable at 14\% 
doping, exactly where the optimal superconducting $T_c$ is observed in 
the cuprates and where the pseudogap is observed to disappear. There is 
a slight first-order re-entrance of SDW at 14\% when the DDW vanishes, 
indicating an intense struggle for dominance between the two kinds of 
order. Neither the specific transition doping densities nor the order 
parameter functional forms are fit.

Why this sequence of phase transitions takes place is easy to understand 
physically.  Figure \ref{f4} shows the evolution of density of states

\begin{equation}
{\cal D}(E) = \frac{1}{8 \pi^2} \sum_{\pm}
\int_{-\pi}^\pi \int_{-\pi}^\pi
 \delta ( E \pm \tilde{E}_{\bf q} )
\; dq_x dq_y
\label{dos}
\end{equation}

\noindent 
as doping is increased.  The rough equivalence of the two orders at 
half-filling, reflected in equality of their energy gaps, becomes 
unbalanced in favor of DDW when carriers are added because DDW, which 
has a node, allows them to be added at zero energy.  Deeper doping then 
destabilizes DDW because it causes the Fermi surface to contract around 
the lines $q_x = \pm q_y$, where the state's node prevents it from 
extracting condensation energy.  SDW order, which can extract 
condensation energy from this region, is then briefly resurrected, but 
it shortly falls victim to the Fermi surface shrinkage that occurs as 
doping is increased further.

The larger implication of Fig. \ref{f3} is that {\it any} system with a 
moment of $s = 0.25$ and rough balance between SDW and DDW at half 
filling will have phase transitions when doped at densities consistent 
with those observed in the cuprates.

\begin{figure}
\includegraphics[scale=0.4]{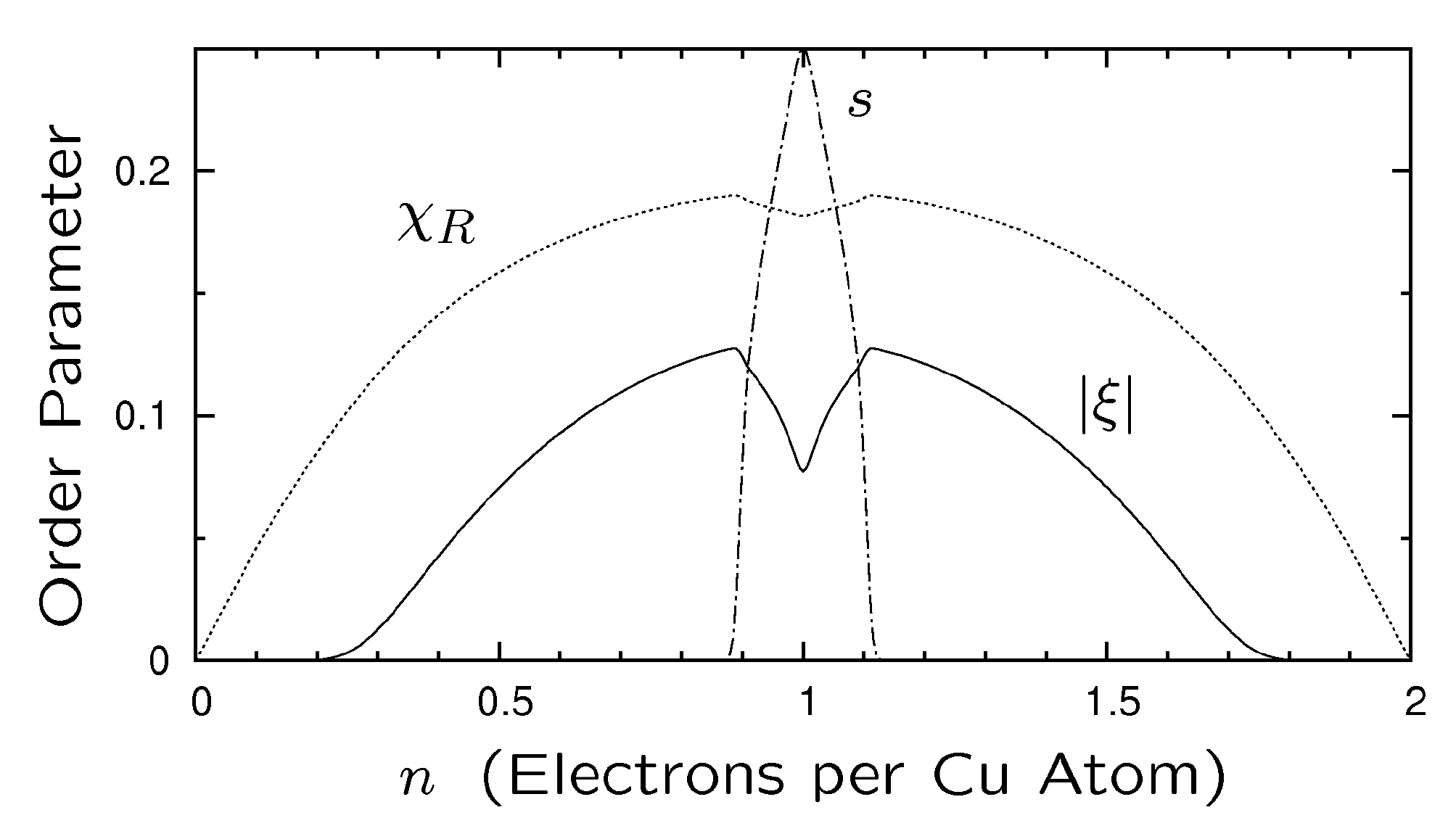}
\caption{Order parameters $s$ and $\xi$ defined by Eqs. (\ref{opdef}) as 
a function of $n$ obtained by solving Eqs. (\ref{scn}) - (\ref{scxi}) 
for the case of $U + 2 J = 5.7 \, t$, $J = 5.9 \, t$, and $t' = V_c = V_n 
= 0$.  This is the same calculation as that in Fig. \ref{f3} but with 
the $\xi = 0$ constraint relaxed. The $n = 1$ value of $|\xi|$ exactly 
equals the $n = 1$ value of $\chi_I$ in Fig. \ref{f3}, as required by
the symmetry described in Appendix \ref{phtrans}.}
\label{f5} 
\end{figure}

\subsection*{SDW vs. DWS}

Let us now repeat the calculation of Fig. \ref{f3} with the $\xi = 0$ 
constraint removed.  The result, shown in Fig. \ref{f5}, reveals that 
DWS now overwhelms DDW completely.  The effect may be understood as an 
analog of a spin flop \cite{demler}. The two kinds of order are exactly 
degenerate at $n = 1$, the way the $x$, $y$ and $z$ components of an 
ideal antiferromagnet are, as described in Appendix \ref{phtrans}. But 
the nesting condition required to stabilize DDW becomes degraded at any 
finite doping while DWS, which does not require nesting, remains robust. 
Adding carriers is thus analogous to adding an anisotropy field to the 
antiferromagnet, and the system responds by ``flopping'' to DWS.

\begin{figure}
\includegraphics[scale=0.4]{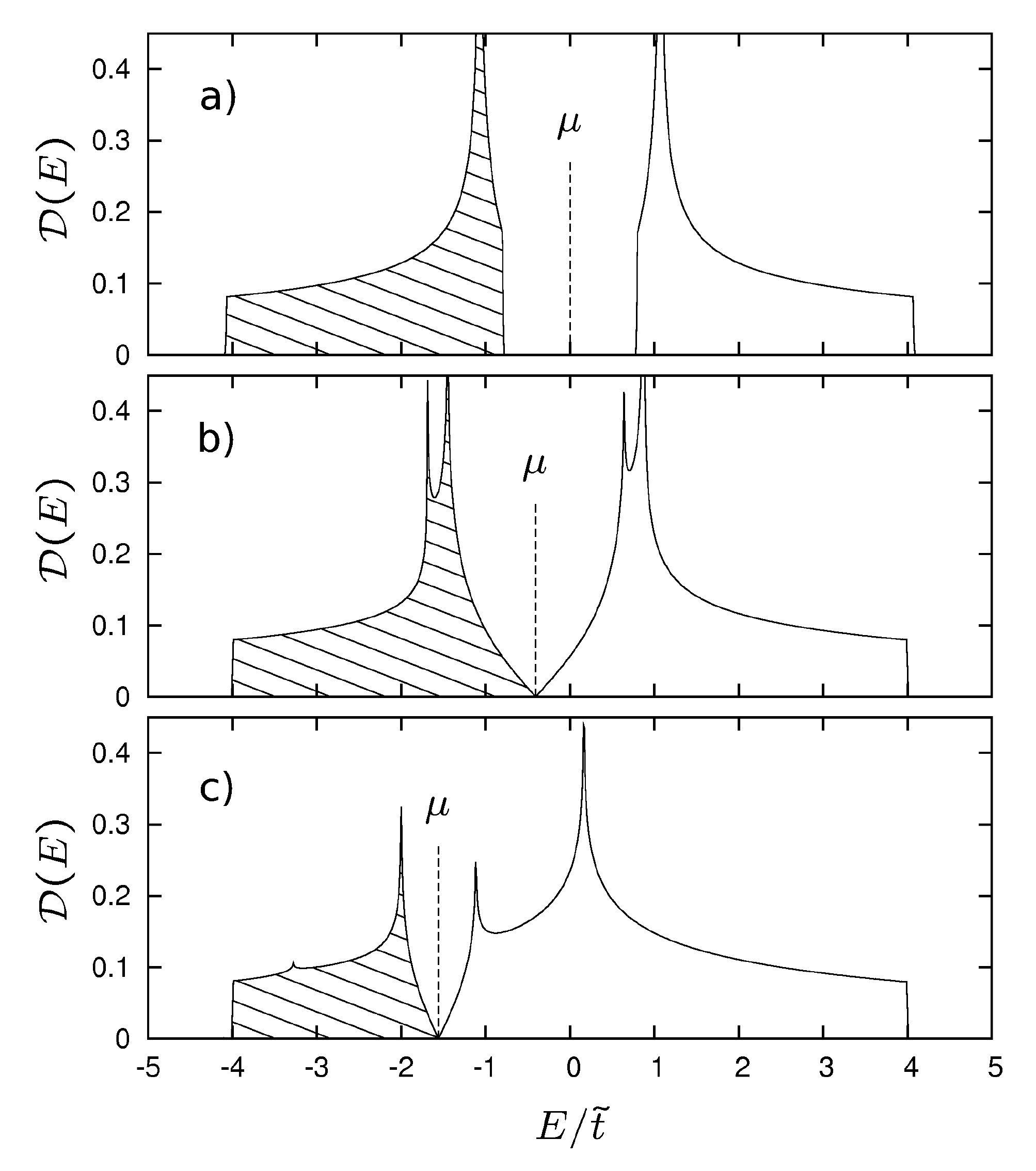}
\caption{Spectral function ${\cal D}(E)$ defined by Eq. (\ref{doss}) 
for the calculation of Fig. \ref{f5} at doping densities (a) $n 0 1.0$, 
(b) $n = 0.865$, and (c) $n = 0.5$.  The system is superconducting in all 
three cases. The energy unit is $\tilde{t}$ defined by Eq. 
(\ref{tildet}).  The shaded region is hole-like (i.e. the signal 
measured in photoemission).}
\label{f6} 
\end{figure}

The persistence of SDW to 12\% doping in Fig. \ref{f5} also shows that 
DWS is less effective at crushing the SDW at low dopings than DDW is, 
even though it is more stable.  The reason is that DWS does not exploit 
any special degeneracies of the band nesting and thus does not use them 
up and make them unavailable to SDW formation the way DDW would.  
Reentrant spin antiferromagnetism is also absent in Fig. \ref{f5}, but 
this is a simple consequence of the persistence of nonzero $\xi$ to high 
dopings.

The coexistence of DWS with SDW everywhere the latter is developed in 
Fig. \ref{f5} is allowed by the special (imposed) symmetries of the 
problem, which guarantee that spin antiferromagnetism fights 
superconductivity through gap formation only, not through pair breaking.  
The system can then become superconducting by exciting electrons across 
the SDW gap quantum mechanically and binding them into Cooper pairs 
there. A system nominally an insulator in this way becomes a 
superconductor. The effect is shown more explicitly in Fig. \ref{f6}, 
where the superconducting spectral function

\begin{displaymath}
{\cal D}(E) = \frac{1}{16 \pi^2} \sum_{\pm}
\end{displaymath}

\begin{displaymath}
\times \int_{-\pi}^\pi \int_{-\pi}^\pi \biggl\{
\biggl[ 1 - \frac{(\pm \tilde{E}_{\bf q} - \tilde{\mu}_{\bf q})}
{E^\pm_{\bf q}} \biggr] \, \delta ( E - \tilde{\mu}_{\bf q}
+ E_{\bf q}^\pm)
\end{displaymath}

\begin{equation}
+ \biggl[ 1 + \frac{(\pm \tilde{E}_{\bf q} - \tilde{\mu}_{\bf q})}
{E^\pm_{\bf q}} \biggr] \, \delta ( E - \tilde{\mu}_{\bf q}
- E_{\bf q}^\pm) \biggr\} \;
dq_x dq_y
\label{doss}
\end{equation}

\noindent
is plotted. As doping increases the SDW gap eventually collapses, 
restoring the $d$-wave node.

\begin{figure}
\includegraphics[scale=0.4]{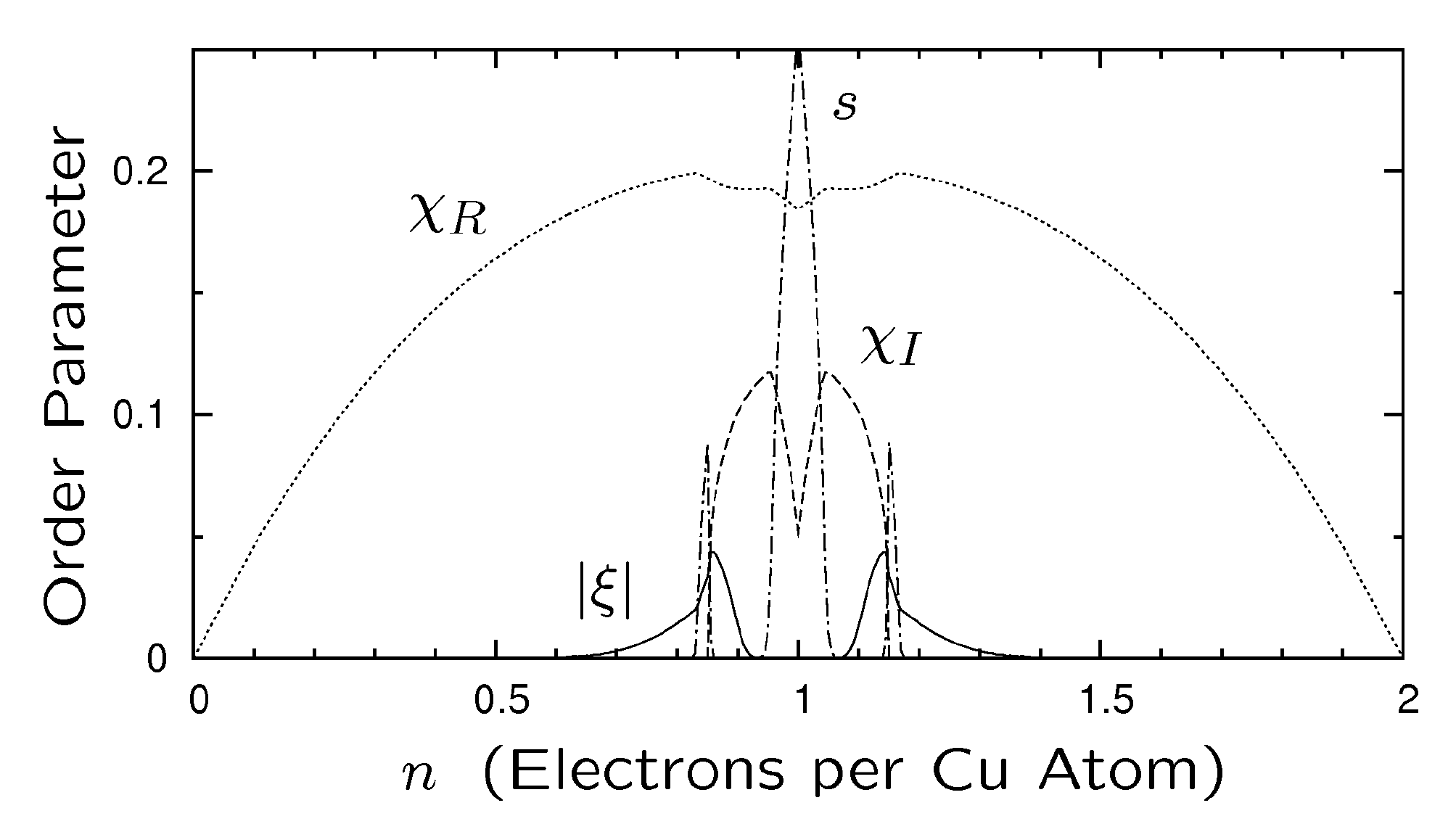}
\caption{Order parameters $s$, $\chi_I$ and $\xi$ defined by Eqs. 
(\ref{opdef}) as a function of $n$ obtained by solving Eqs. (\ref{scn}) 
- (\ref{scxi}) for the case of $U + 2 J = 2.5 \, t$, $V_c = 1.3 \, t$, 
$J = 0.7 \, t$, and $t' = V_n = V_t =0$.  The parameters $(U + 2 
J)/\tilde{t}$ and $(3 J/4 + V_n + V_c)/\tilde{t}$, with $\tilde{t}$ 
defined as in Eq. (\ref{tildet}), are about the same as those assumed
in Figs. \ref{f3} and \ref{f5}. The reentrant spin antiferromagnetism 
at 15\% doping is the same as that seen in Fig. \ref{f3}.}
\label{f7} 
\end{figure}

\subsection*{DWS vs. DDW}

Let us now get all three order parameters to appear in the phase diagram 
by repeating the calculation of Fig. \ref{f5} in the presence of $V_c 
>0$. This parameter breaks the degeneracy between DDW and DWS at half 
filling, encouraging the former over the latter. The result is shown in 
Fig. \ref{f7}.  In order to maintain the half-filling conditions 
implicit in Figs. \ref{f3} and \ref{f5}, we accompany the increase of 
$V_c$ with an adjustment of the parameters $U$ and $J$ that keeps $(U + 
2 J)/\tilde{t}$ and $(3 J/4 + V_n + V_c)/\tilde{t}$ constant, per Eqs. 
(\ref{tildet}) and (\ref{scs}) - (\ref{scci}). When $V_c$ is increased 
to the value used in Fig. \ref{f7}, this causes the interaction 
parameters $U$, $J$ and $V_c$ all to become reasonably sized (i.e. 
comparable to $t$), and, even more importantly, causes $U$ to switch 
from negative to positive. The behavior of Fig. \ref{f3} is now 
restored, this time legitimately, but DDW is accompanied by 
doping-dependent DWS, with which it coexists.  The latter first acquires 
significant magnitude when the SDW is destroyed at 5\% and then rises up 
as the DDW dies away, peaking at 15\% where the latter disappears, and 
then declining rapidly.  The sequence of events is identical to that 
observed in $p$-type cuprates.

The result of Fig. \ref{f7} cannot be achieved using $V_n > 0$. The 
half-filling degeneracy of DDW and DWS is broken the same way by both 
parameters, but $V_n > 0$ has the effect of suppressing DWS, per Eq. 
(\ref{scxi}).  This suppression can be counteracted by increasing the 
value of $J$, but this then requires making $U$ more negative, per Eq. 
(\ref{scs}). Negative $U$ is highly unphysical. Were it present, it 
would stabilize $s$-wave superconductivity, a phenomenon not observed in 
the cuprates.

\subsection*{Particle-Hole Asymmetry} 

\begin{figure}
\includegraphics[scale=0.4]{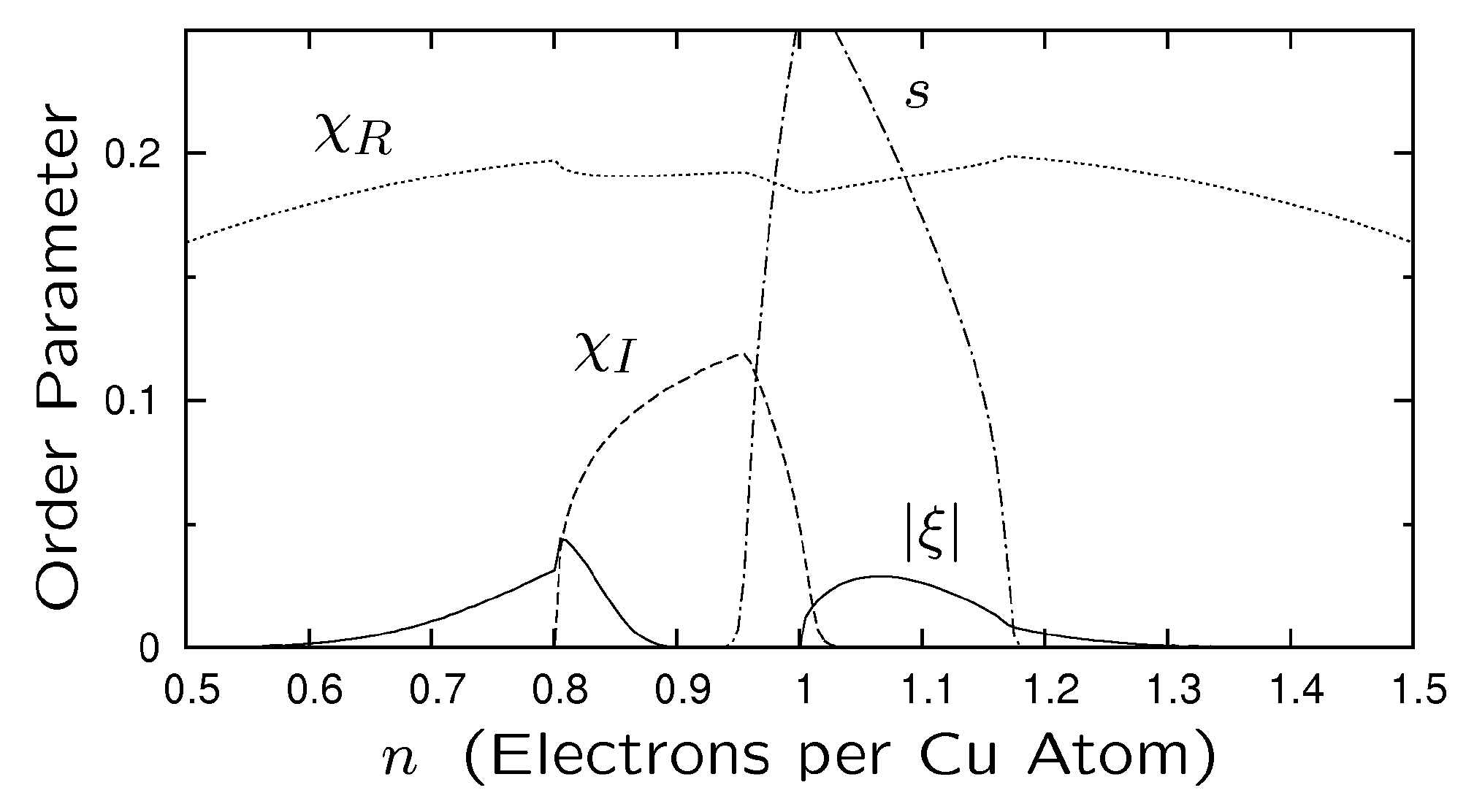}
\caption{The same calculation as in Fig. \ref{f7} except with band 
asymmetry $t' = 0.1 \, t$. The remaining parameters are $U + 2 J = 2.5 
\, t$, $V_c = 1.3 \, t$ $J = 0.7 \, t$, and $V_n = V_t = 0$.
The large region of coexistence between DWS and SDW on the $n$-type side 
is an artifact of having constrained the SDW to be periodic in the 
doubled unit cell.  When this condition is relaxed, the system insulates 
due to domain wall trapping everywhere SDW order is developed
\cite{schulz,inui,sarkar,chubukov,gong}}.
\label{f8} 
\end{figure}

Let us now introduce particle-hole asymmetry by repeating the 
calculation of Fig. \ref{f7} with an added $t' = 0.1 \, t$. The result is 
shown in Fig. \ref{f8}.  The pseudogap is now completely absent on the 
$n$-type side, and the re-entrant spin antiferromagnetism has disappeared 
from both sides.

The ability of such a small change in the underlying band structure to 
violently rearrange the phase diagram is consistent with the variability 
of cuprate experiments, both among different materials and among samples 
of the same material prepared different ways.  It occurs because small 
perturbations tip the fine balance among orderings.  Defects and crystal 
boundaries would be expected to rearrange the order parameters locally 
in a similar fashion, thus causing large effects that have no analog in 
semiconductors. This result is also consistent with the observed 
complexity of lattice instabilities in these materials \cite{tranquada}. 
Describing the latter requires addition of an electron-phonon 
interaction to ${\cal H}_0 + \Delta {\cal H}$, but doing so is 
straightforward given that the only effect of moving atoms around is to 
change the underlying band structure.

\section{Parametric Constraints}

\label{oxygen}

Figures \ref{f7} and \ref{f8} show that the interaction parameters $J$, 
$U$, and $V_c$ are fixed by the magnitudes of $s$, $\chi_I$, and $\xi$, 
and that $V_n = 0$ is fixed by the overall shape of the phase diagram. 
The conclusion that $V_n = 0$ resolves the important controversy of 
whether antiferromagnetic exchange or near-neighbor attraction, possibly 
phonon-mediated, causes high-$T_c$ superconductivity.  Only the former 
is both physically reasonable and compatible with experiment. 

The reason that one cannot have $V_n < 0$ is subtle and requires some 
discussion. Only $J > 0$ and $V_n < 0$ have the ability to stabilize 
$d$-wave superconductivity, so at least one of them must be present and 
sizable. However, a $J > 0$ of the requisite magnitude is already 
present, as demonstrated by the antiferromagnetism of the undoped 
insulator. Adding $V_n < 0$ would further stabilize $d$-wave 
superconductivity, but unfortunately it would do so near half-filling 
and destabilize the pseudogap.  To restore the pseudogap, one would then 
have to compensate by making either $J$ or $V_c$ more positive, per Eq. 
({\ref{scci}). But, $J$ actually has to {\it decrease} if the 
superconducting aspects of Fig. \ref{f7} are to remain the same, per 
Eq. (\ref{scxi}).  The required increase in $V_c$ would then severely 
narrow the quasiparticle bandwidth, per Eq. (\ref{tildet}), an effect 
not observed experimentally.  Accordingly, $V_n$ can be neither negative 
nor positive but must be zero.

The conclusion that $V_n = 0$ also makes sense physically.  The 
parameter $V_n$ is a Coulomb interaction. There is fundamentally no 
reason for it to be negative, just as there is no reason for $U$ to be 
negative. Indeed one's first guess would be that both parameters had 
been mostly accounted for in the generation of the band structure.  A 
parameter $V_n < 0$ would also tend to stabilize $s$-wave 
superconductivity unless prevented from doing so by a sufficiently large 
$U$.  The latter would have to be large enough to prevent mixed $d + s$ 
superconductivity, a phenomenon not observed in any part of the cuprate 
phase diagram.  The parameter $J > 0$ by contrast is not only 
demonstrably present but something unique to the cuprates.

In addition to properly balancing the phase diagram, the parameter $V_c 
> 0$ has three important effects that indicate it is not only useful
but actually physically necessary: (1) it enables $U$ to be positive, 
(2) it reduces $J$ to a reasonable size, and (3) it enables $\tilde{t}$ 
to be less than $t$. The latter is particularly important. The parameter 
$\tilde{t}$ fixes physical quasiparticle bandwidth, a quantity observed 
experimentally not to be broadened, but it also sets quasiparticle Fermi 
velocity.  Were $\tilde{t}/t > 1$, the oscillator strength of the Fermi 
sea conductivity pole would exceed the total $f$-sum rule, which is 
fixed by $t$ solely, thus indicating that the system was not in its 
ground state.  This sum rule problem is discussed further Section 
\ref{superprops} in the context of the superfluid density.

\begin{figure}
\includegraphics[scale=0.4]{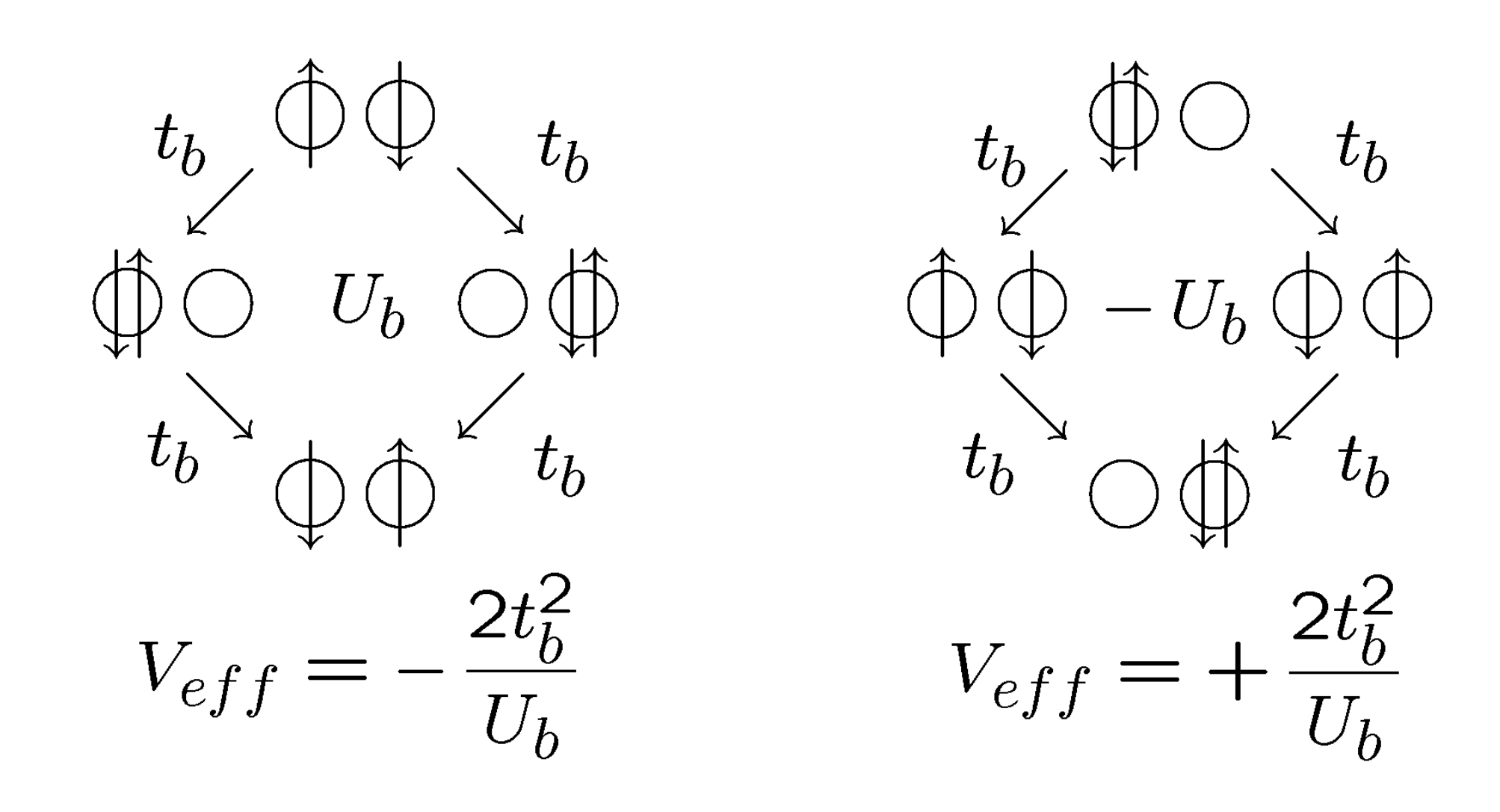}
\caption{Illustration of the generation of $V_c$ and $J$ together 
through superexchange as described by Eq. (\ref{bondham}).  Cuprate 
antiferromagnetism is caused by blocking. \cite{superex} The bond 
quantum mechanics thus deforms adiabatically to a Hubbard model 
characterized by parameters $t_b$ and $U_b$. Second-order perturbation 
theory then gives an effective matrix element $V_{eff}$ for exchanging a 
pair of electrons that is negative because the intermediate states have 
positive energy. The matrix element for Cooper pair tunneling has the 
opposite sign because the intermediate states have negative energy.}
\label{f9} 
\end{figure}

Given the insight that postulating $J > 0$ makes no sense without also 
postulating $V_c > 0$ of roughly the same magnitude, the source of the 
latter is easily identified: It is a secondary aspect of superexchange. 
This is illustrated in Fig. \ref{f9}. Spin antiferromagnetism is due to 
blocking \cite{superex}. One knows this because the spin-orbit and 
magnetic dipole interactions in typical antiferromagnets are too small 
to account for the observed spin stiffness.  The bond quantum mechanics 
may therefore be adiabatically deformed to a Hubbard model characterized 
by parameters $t_b$ and $U_b$. In terms of the four configurational 
states

\begin{displaymath}
| 1 \! > =
c_{j \uparrow}^\dagger c_{j \downarrow}^\dagger | 0 \! >
\; \; \; \; \; 
| 2 \! > =
c_{k \uparrow}^\dagger c_{k \downarrow}^\dagger | 0 \! >
\end{displaymath}

\begin{equation}
| 3 \! > =
c_{j \uparrow}^\dagger c_{k \downarrow}^\dagger | 0 \! >
\; \; \; \; \; 
| 4 \! > =
c_{k \uparrow}^\dagger c_{j \downarrow}^\dagger | 0 \! >
\end{equation}

\noindent
we then have the Hamiltonian

\begin{equation}
{\cal H}_{bond} = \left[ \begin{array}{cccc}
U_b & 0 & t_b & t_b \\
0 & U_b & t_b & t_b \\
t_b & t_b & 0 & 0 \\
t_b & t_b & 0 & 0 \\ \end{array} \right]
\label{bondham}
\end{equation}

\noindent
Second-order perturbation theory then gives a spin-exchange matrix
element of $- 2t_b^2/U_b$ and a Cooper pair tunneling matrix element
of $+ 2t_b^2/U_b$.

Correlation corrections to the properties of metals are notoriously 
difficult to calculate from first principles, especially if they are 
performed by summing Feynman graphs.  In graphical sums, the 
negative-energy denominators required to obtain $V_c > 0$ show up in the 
reverse time orderings \cite{fetter}. Such calculations are beyond the 
scope of this work. The purpose of Eq. (\ref{bondham}) is only to show 
that it is physically reasonable for $V_c > 0$ to appear whenever $J > 
0$ does.

The parameter fits used in generating Figs. \ref{f7} and \ref{f8} thus 
lead to the conclusion that superexchange mediated by the bonding oxygen 
atom, not Coulomb repulsion on the copper sites, is responsible for all 
three ordering phenomena.

\begin{figure}
\includegraphics[scale=0.4]{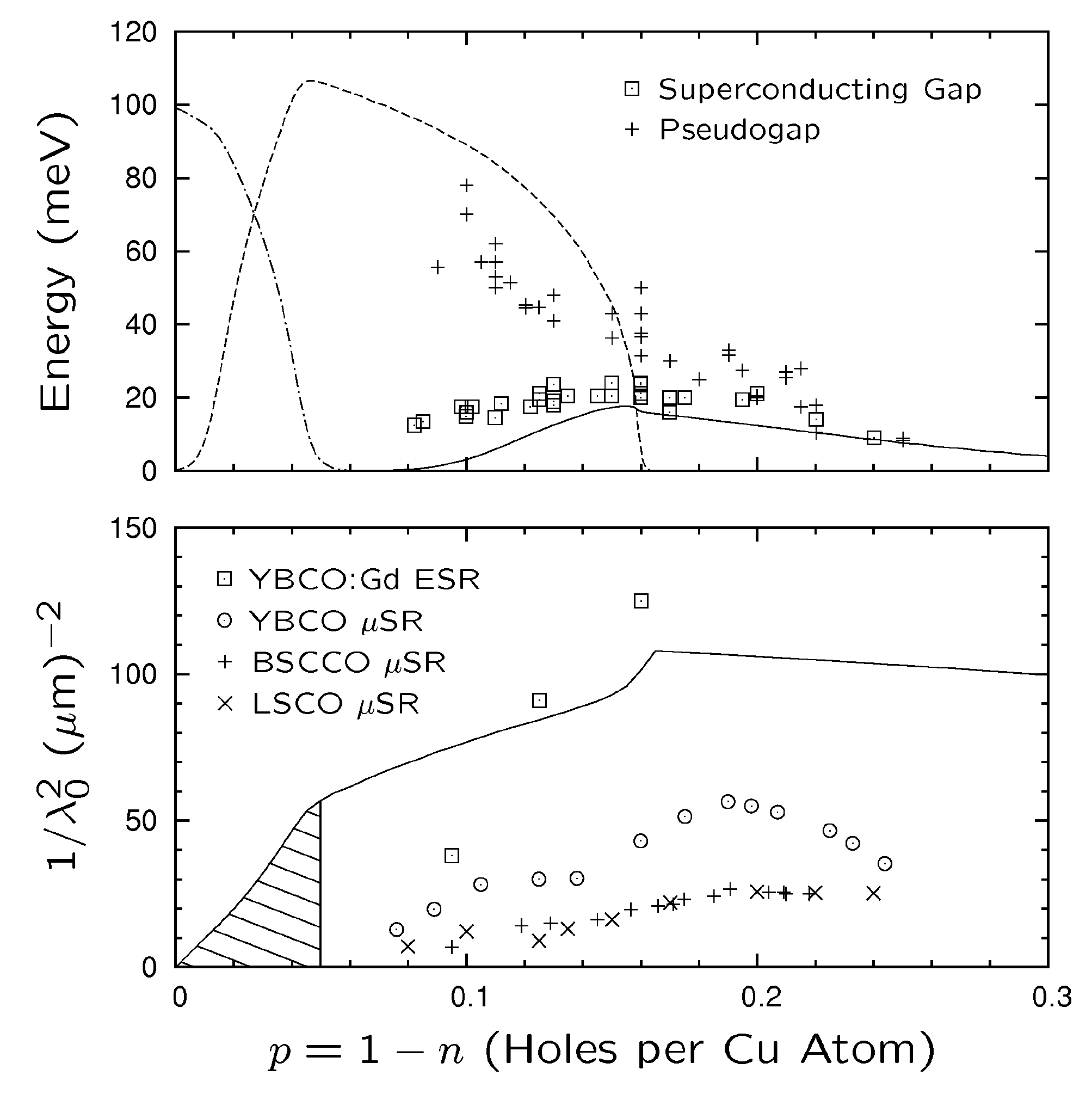}
\caption{Top: Comparison of SDW gap $(U + 2 J) s$ (dash-dotted line), maximum 
DDW gap $(3 J + 4 V_c) \chi_I$ (dashed line) and maximum superconducting gap 
$3 J |\xi|$ (solid line) computed using Eqs. (\ref{scn}) - (\ref{scxi}) with 
values of the pseudogap ($+$) and superconducting gaps ($\Box$) inferred 
from a variety of experiments by H\"{u}fner {\it et al.} \cite{hufner}. 
Bottom: London penetration depth calculated with Eq. (\ref{london}) 
compared with $\mu$SR measurements on polycrystalline samples reported 
by Tallon {\it et al} ($+$, $\times$, $\circ$) and in-plane ESR 
measurements on oxygen-ordered YBCO:Gd samples by Pereg-Barnea {\it et 
al.} \cite{tallon,bernhard,pereg}. About half the disparity in the YBCO 
results are attributable to penetration depth anisotropy, which causes 
the powder average of $1/\lambda_0^2$ to be about 2/3 of the in-plane 
one. The hatching indicates insulation due to SDW domain-wall trapping. 
The oscillator strength in this region does not reside at zero frequency 
but instead at a low, but finite, frequency characteristic of the 
trapping \cite{uchida,padilla}. The calculations assume the parameters 
of Fig. \ref{f1}: $U = 0.76 \, t$, $J = 0.75 \, t$, $V_c = 0.87 \, t$, 
$t' = 0.1 \, t$, $V_n = V_t = 0$, and $t = 0.19$ eV.}
\label{f10} 
\end{figure}

\section{Superconducting Properties}

\label{superprops}

In what follows we shall assume the parameters used to generate Fig. 
\ref{f1}: $U = 0.76 \, t$, $J = 0.75 \, t$, $V_c = 0.87 \, t$, $t' = 0.1 
\, t$, $V_n = V_t = 0$, and $t = 0.19$ eV.  These amount to a small 
fine-tuning of the parameters of Fig. \ref{f8}.

\subsection*{Energy Gap}

Figure \ref{f10} compares the maximum gaps computed using Eq. (\ref{scn}) 
- (\ref{scxi}) with pseudogap and superconducting gap value estimated by 
H\"{u}fner {\it et al} from a variety of experimental sources 
\cite{hufner}. The latter include scanning tunneling microscopy, 
photoemission, Raman scattering, break junction tunneling, magnetic 
resonance, inelastic neutron scattering, thermal transport and Andreev 
scattering 
\cite{ding,peets,venturini,sugai,zasadzinski,sidis,he,hawthorn,sutherland,deutscher}. 
The experimental plot is qualitatively similar to that of Le Tacon {\it 
et al.} and of Valenzuela and Bascones but is in absolute units 
\cite{tacon,valenzuela}. The assignment of specific values to the 
experimental gaps is somewhat subjective because they are not sharply 
defined. A reasonable estimate of the error bar is 30\%.  Particularly 
important are the Andreev reflection experiments, which show that 
particle-hole mixing characteristic of superconducting order vanishes at 
an energy scale much lower than that of the pseudogap 
\cite{deutscher,gonnelli}.

Figure \ref{f10} severely constrains the choice of $U$, $J$, and $V_c$. The 
scale of superconducting gap fixes $J = 0.75 \, t$. The scale of the 
pseudogap fixes $3 J/4 + V_c = 1.43 \, t$.  The condition that $s \simeq 
0.25$ at half filling fixes $U + 2 J = 2.26 \, t$. These parameters then 
place the system at the edge of the half-filling coexistence region of 
Fig. \ref{f2} (with the substitution $J \rightarrow J + 4 V_c/3$ keeping 
$U + 2J$ fixed), thus causing the two phase transitions take place at 
their experimentally observed doping values of 5\% and 16\%.

\subsection*{London Penetration Depth}

The zero-temperature London penetration depth $\lambda_0$ is given by

\begin{equation}
\frac{1}{\lambda_0^2} = 
\frac{4 \pi e^2}{\hbar^2 c^2} (\frac{t}{a}) \, \tilde{n}_s
= 151 \; (\mu {\rm m})^{-2} \; \tilde{n}_s
\label{london}
\end{equation}

\noindent
where $a = 5.84$ \AA $\,$ is an interlayer spacing appropriate
for YBCO and $\tilde{n}_s$ in the effective superfluid density per Cu atom

\begin{displaymath}
\tilde{n}_s = \frac{1}{16 \pi^2 t} \sum_\pm
\end{displaymath}

\begin{equation}
\int_{-\pi}^\pi \int_{-\pi}^\pi
\biggl[ 1 - \frac{(\pm \tilde{E}_{\bf q} - \tilde{\mu})}
{E_{\bf q}^\pm} \biggr] \nabla_q^2 (\pm \tilde{E}_{\bf q} - \tilde{\mu})
\label{sdense}
\end{equation}

\noindent
The superfluid density is not a well-defined quantity, so the choice of 
$t$ as the conversion factor between $1/\lambda_0^2$ and $\tilde{n}_s$ is 
somewhat arbitrary.  It corresponds to the effective mass formula

\begin{equation}
\frac{1}{\lambda_0^2} = 
\frac{4 \pi e^2}{m^* c^2} (\frac{\tilde{n}_s}{a b^2})
\; \; \; \; \; \; \; 
m^* = \frac{\hbar^2}{t \, b^2}
= 2.50 \, m_e
\end{equation}

\noindent
with a bond length $b = 4$ \AA. 

The formal justification of Eq. (\ref{sdense}) when DDW order is 
developed is complicated by the fact that the DDW order parameter 
$\chi_I$ is gauge-variant, while the Hamiltonian parameter $J$ that 
gives rise to it is not.  Handling this properly requires executing a 
vertex correction, which is technically beyond the scope of this work. 
However, it is easy to see on physical grounds that Eq. (\ref{sdense}) 
must be correct. It says that each quasiparticle carries electric charge 
$e$ and moves with a speed that is the momentum derivative of its 
energy, just as occurs in SDW.  The quasiparticles of DDW and SDW must 
behave similarly because orbital antiferromagnetism and spin 
antiferromagnetism are aptly analogous.

Figure \ref{f11} shows the superfluid density $\tilde{n}_s$ calculated 
using Eq. (\ref{sdense}) and the parameters of Fig. \ref{f1} as a 
function of doping.  It is free-electron-like except when the DDW and 
SDW orders develop near $n \rightarrow 1$, when it is strongly 
suppressed.  This reflects the reorganization of the Fermi surface into 
pockets. Loss of superfluid density due to an elementary Fermi surface 
reconstruction is fully consistent with the finding of Tallon {\it et 
al.} that the ratio of the superconducting temperature $T_c$ and the 
specific heat jump remains constant in this limit while both quantities 
individually vanish \cite{tallon}.

The values of $1/\lambda_0^2$ implicit in Fig. \ref{f11} through Eq. 
(\ref{london}) are compared with experiment in Fig. \ref{f10} 
\cite{tallon,bernhard,pereg}. The large variability among them is 
symptomatic of disorder degradation.  The proper comparison to make is 
thus with the best samples at optimal doping.  The good agreement in 
these cases confirms $\tilde{n}_s \simeq 1$.

\begin{figure}
\includegraphics[scale=0.4]{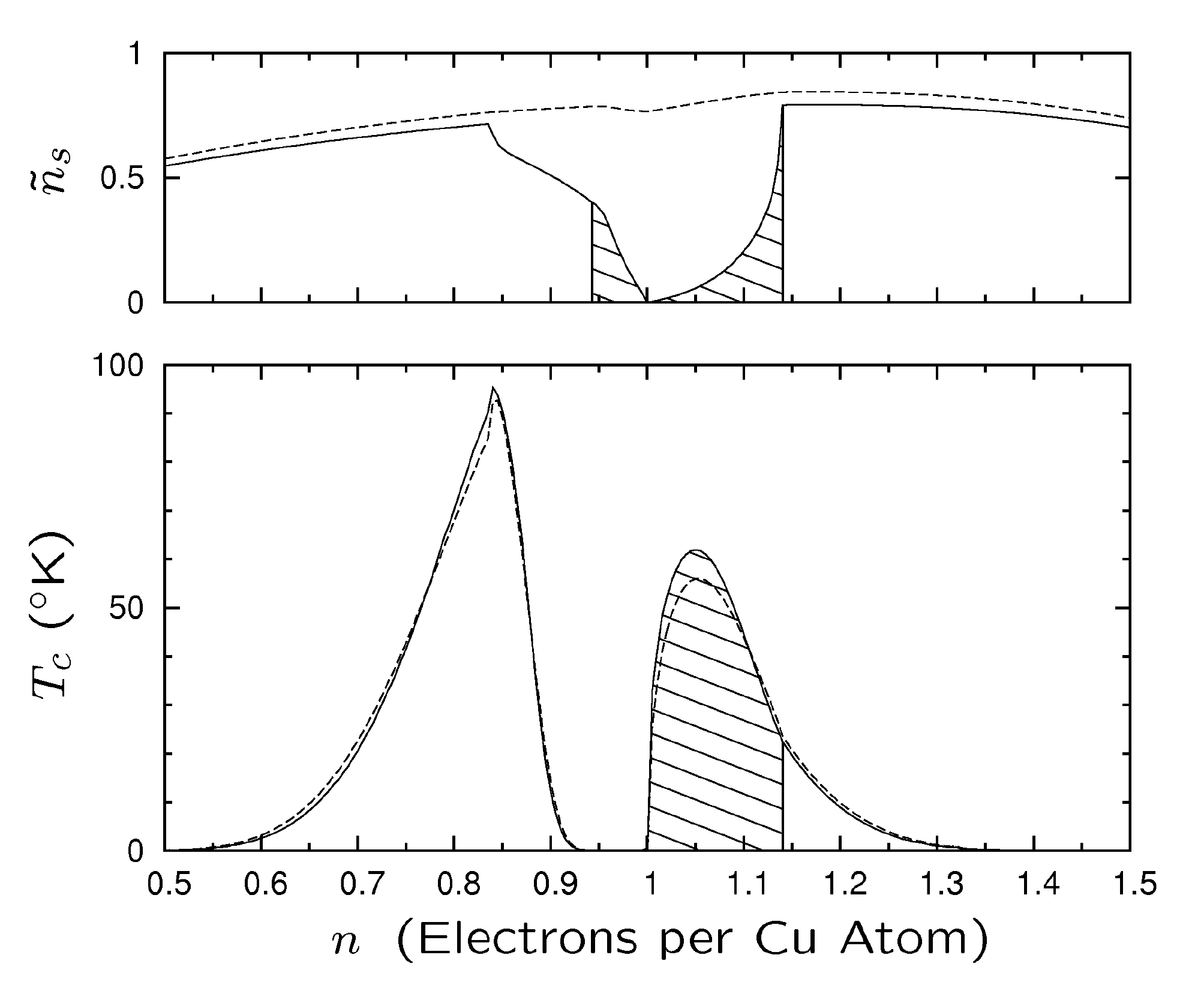}
\caption{Top: Superfluid density $\tilde{n}_s$ defined by Eq. 
(\ref{sdense}).  The dashed curve is the total f-sum rule given by
Eq. (\ref{fsum}).  Bottom: Superconducting transition temperature
defined by Eq. (\ref{tc}).  The dashed curve is the maximum superconducting
gap divided by 2.2, i.e. $3 J |\xi| /2.2 k_B$. The hatching
has the same meaning as in Fig. \ref{f10}: This region
is insulating, rather than superconducting, because of carrier trapping
at SDW domain walls. The computations assume the parameters of Fig. 
\ref{f1}:  $U = 0.76 \, t$, $J = 0.75 \, t$, $V_c =0.87 \, t$, 
$t' = 0.1 \, t$, $V_n = V_t = 0$, and $t = 0.19$ eV.}
\label{f11} 
\end{figure}

The nonzero superfluid density for $0.95 < n < 1.14$ in Fig. \ref{f11} 
($0 < p < 0.05$ in Fig. \ref{f10}) is an artifact of having forced the 
SDW to periodic in the doubled unit cell.  Relaxing this condition will 
cause the quasiparticles to trap in domain walls, thus forcing the 
superfluid density to zero, in agreement with experiment 
\cite{schulz,inui,sarkar,chubukov,gong}. The system is an insulator over 
the entire range in which SDW is developed---with the possible exception 
of $n \simeq 1.4$, where coexistence of SDW and DWS has been reported 
\cite{yu,luo}. The superfluid oscillator strength in question, however, 
is not lost but is simply transported upward to a small, but finite 
frequency characteristic of the trapping. This is seen experimentally in 
the lightly-doped cuprates as mid-infrared absorption 
\cite{uchida,padilla}. It is centered at about 0.6 eV in 
La$_{2-x}$Sr$_x$CuO$_4$.

Figure \ref{f11} also shows that the total $f$-sum rule

\begin{displaymath}
\tilde{n}_{total} = 4 \chi_R
- 8 \, (\frac{t'}{t}) \,\chi_R'
\end{displaymath}

\noindent
derived in Appendix \ref{fsumrule}, where

\begin{displaymath}
\chi_R' = \frac{1}{64 \pi^2 t'} \sum_\pm
\end{displaymath}

\begin{equation}
\times \int_{-\pi}^\pi \int_{-\pi}^\pi
\biggl[ 1 - \frac{(\pm \tilde{E}_{\bf q} - \tilde{\mu})}
{E_{\bf q}^\pm} \biggr] \, (\mu - \tilde{\mu})
\; dq_x dq_y
\label{fsum}
\end{equation}

\noindent
per Eqs. (\ref{opdef}), exceeds $\tilde{n}_s$.  This is an important 
stability test.  $\tilde{n}_{total}$ is the sum of $\tilde{n}_s$ plus 
all the additional optical oscillator strength at higher frequencies.  
If one had $\tilde{n}_s > \tilde{n}_{total}$ at any doping value, it 
would imply that the system had negative oscillator strength (laser 
gain) at higher frequencies, and thus that it was not in its ground 
state.  But Eqs. (\ref{scn}) - (\ref{scxi}) show that $\tilde{n}_s < 
\tilde{n}_{total}$ is enforced by $(3 J/4 - V_c) < 0$.  Thus Fig. 
\ref{f11} also shows that the presence of a $V_c > 0$ is required for $J
> 0$ to make sense physically. Moreover, the particular values of these 
parameters used to generate Fig. \ref{f11} give $\tilde{n}_s$ comparable 
to $\tilde{n}_{total}$ except where the fermi surface is reconstructed. 
This means that nearly all of the oscillator strength is exhausted by 
the fermi surface pole, as would be expected on physical grounds.

\subsection*{Transition Temperature $T_c$}

The relative largeness of the superfluid density near optimal doping 
implies that the superconducting transition temperature is not 
determined by phase fluctuations \cite{emery}. The phase decoherence 
temperature cannot be any lower than the Kosterlitz-Thouless temperature 
\cite{benfatto}

\begin{equation}
k_B T_{KT} = \frac{\pi}{8} \frac{\hbar^2}{m^* b^2} \, 
\tilde{n}_s = \frac{\pi}{8} t \, \tilde{n}_s
\end{equation}

\noindent
which lies above 433 $^\circ$K ($\tilde{n}_s = 0.5$) over the entire
range of interest. It is thus too high to matter.

The transition temperature must therefore be determined by the weakening
of the DWS order parameter by excitation of conventional BCS
quasiparticles.  It is computed by substituting $\xi = 0$ in
Eqs. (\ref{scn}) - (\ref{scxi}) and then replacing Eq. (\ref{scxi}) with

\begin{displaymath}
1 = (\frac{3}{4} J) \, \frac{1}{16 \pi^2} \sum_\pm
\int_{-\pi}^\pi  \int_{-\pi}^\pi
\frac{1}{|\pm \tilde{E}_{\bf q} - \tilde{\mu} |} \, 
\end{displaymath}

\begin{equation}
\times 
\tanh ( \frac{E_{\bf q}^\pm}{2 k_B T_c} ) \,
\biggl[ \cos(q_x) - \cos(q_y) \biggr]^2 \,
dq_x dq_y
\label{tc}
\end{equation}

\noindent
One obtains the transition temperatures shown in Fig. \ref{f11}.
Also shown is the maximum $T = 0$ superconducting gap $4 J |\xi|$
divided by $2.2 k_B$.  The factor of 2.2 is similar to the 2.37
reported by Tallon {\it et al.} for a slightly different model
\cite{tallon}.  A factor of approximately 2.5 was assumed by
H\"{u}fner {\it et al.} in inferring the data points plotted in
Fig. \ref{f10} \cite{hufner}.

\section{Other Properties}

\label{otherprops}

\subsection*{Insulation}

The SDW equations are well known to be unstable to antiferromagnetic 
domain wall formation whenever the SDW gap is nonzero 
\cite{schulz,inui,sarkar,chubukov,gong}. This is consistent with the 
antiferromagnetic discommensuration observed by neutron scattering in 
doped insulating La$_{1.95}$Sr$_{0.05}$CuO$_4$ \cite{wakimoto}. The 
mathematics of twisting is briefly reviewed in Appendix 
\ref{spiraltrap}. For the parameter range considered here, domain wall 
formation has no significant effect on the SDW gap magnitude but simply 
causes the added carriers to trap. This trapping occurs in two ways.  
The first is that domain walls already present preferentially bind added 
carriers the way impurities in a semiconductor do. The second is that 
added carriers create domain walls and self-trap in them if the system 
is annealed. Domain wall formation is inherently glassy and thus 
difficult to describe mathematically. It is also sensitive to details, 
such as sample preparation and the 5 meV anisotropy energy associated 
with preferential ordering of the spins in the $x$-$y$ plane \cite{keimer}. 
The latter is due to spin-orbit coupling, which is left out of Eq. 
(\ref{dh}). As a consequence, there is presently no general agreement on 
what the true ground state of the doped antiferromagnet is.

The small region of coexistence between SDW and DWS in the range $1.12 < 
n < 1.15$ reported by Yu {\it et al.} and Luo {\it et al.} is consistent 
with Fig. \ref{f11} \cite{yu,luo}. The SDW and DWS order parameters lie 
in different irreducible representations of the system's symmetry group. 
Provided that they are competing energetically and that a conventional 
Ginzburg-Landau description remains correct in the presence of disorder, 
the system must have either (1) a first-order transition between them or 
(2) two second-order transitions in and out of a coexistence region.

The trapping effects in the cuprates actually extend over the entire 
insulating range $0.95 < n < 1.14$.  This must be so because insulation 
is physically impossible in a translationally-invariant system that can 
be continuously doped. The proof is very simple: Applying a nonuniform 
voltage is the same thing as changing the chemical potential locally. 
Thus the conventional metallic behavior in the range $0.95 < n < 1$ 
implicit in Fig. \ref{f11} is also an artifact.

Domain wall formation is also an issue with the DDW order, which is 
arguably the cause of the glassiness observed in scanning tunneling 
microscopy near optimal doping \cite{mcelroy}. However it is less 
serious in that case because (1) the DDW state has a node at which 
carriers may be added with minimal energy and (2) the DDW order 
parameter is Ising-like.  The latter implies that continuous twisting is 
impossible, and thus that domain walls have a much higher energy cost 
than the domain walls of the SDW do.

\subsection*{Magnon Spectrum}

The ladder approximation for the spin susceptibility is given by 
\cite{frenkel}

\begin{displaymath}
\chi_{\bf q}(\omega) = \biggl\{ 
[\chi_{\bf q}^{(0)}(\omega) ]^{-1} + U \, \left[
\begin{array}{cc} 1 & 0 \\
0 & 1 \\ \end{array} \right]
\end{displaymath}

\begin{equation}
- J [ \cos(q_x) + \cos(q_y)]
 \left[ \begin{array}{cc} 1 & 0 \\
0 & - 1 \\ \end{array} \right] 
\biggr\}^{-1}
\label{rpa}
\end{equation}

\noindent
where

\begin{displaymath}
\chi_{\bf q}^{(0)} (\omega)
= \frac{1}{8 \pi^2} \int_{\pi}^\pi \int_{\pi}^\pi
\biggl\{ \frac{1}{2 E_{{\bf k} + {\bf q}} E_{\bf k}}
\end{displaymath}

\begin{displaymath}
\times \left[
\begin{array}{cc}
E_{{\bf k} + {\bf q}} E_{\bf k} - \tilde{\epsilon}_{{\bf k} 
+ {\bf q}} \tilde{\epsilon}_{\bf k} + \tilde{\Delta}^2 &
\tilde{\Delta} (E_{{\bf k} + {\bf q}} + E_{\bf k}) \\
\tilde{\Delta} (E_{{\bf k} + {\bf q}} + E_{\bf k}) &
E_{{\bf k} + {\bf q}} E_{\bf k} - \tilde{\epsilon}_{{\bf k} 
+ {\bf q}} \tilde{\epsilon}_{\bf k} + \tilde{\Delta}^2 \\
\end{array} \right]
\end{displaymath}

\begin{displaymath}
\times \frac{1}{\hbar \omega - (E_{{\bf k} + {\bf q}} +
E_{\bf k} ) + i \eta} + \frac{1}{2 E_{{\bf k} + {\bf q}} E_{\bf k}}
\end{displaymath}

\begin{displaymath}
\times \left[
\begin{array}{cc}
E_{{\bf k} + {\bf q}} E_{\bf k} - \tilde{\epsilon}_{{\bf k} 
+ {\bf q}} \tilde{\epsilon}_{\bf k} + \tilde{\Delta}^2 &
- \tilde{\Delta} (E_{{\bf k} + {\bf q}} + E_{\bf k}) \\
- \tilde{\Delta} (E_{{\bf k} + {\bf q}} + E_{\bf k}) &
E_{{\bf k} + {\bf q}} E_{\bf k} - \tilde{\epsilon}_{{\bf k} 
+ {\bf q}} \tilde{\epsilon}_{\bf k} + \tilde{\Delta}^2 \\
\end{array} \right]
\end{displaymath}

\begin{equation}
\times \frac{1}{- \hbar \omega - (E_{{\bf k} + {\bf q}} +
E_{\bf k} ) - i \eta} \biggr\} \, dq_x dq_y
\label{chi0}
\end{equation}

\noindent 
The ladder sum is formally similar to the random phase approximation in 
dielectric response, and is often referred to as the RPA for this reason 
even though the graphs in question are of the opposite sign. Including 
it is a standard method of computing spin fluctuation properties of SDW 
states \cite{nakanishi}. Figure \ref{f12} shows how it effectively adds 
a large attractive interaction between the excited particle and hole, 
thereby binding them down out of the continuum to form a spin-1 magnon. 
The spin wave dispersion curve that results is compared in Fig. 
\ref{f12} with the inelastic neutron scattering measurements on 
La$_2$CuO$_4$ reported by Coldea {\it et al.} \cite{coldea}. Although the 
RPA is notoriously inaccurate at short wavelengths, the disparity at the 
Brillouin zone edge in Fig. \ref{f12} is most likely not a computational 
problem but an SDW gap that is too small, particularly since this 
material has a larger magnetic moment than the other undoped cuprates 
\cite{tranquada}. Raising $U$ to $1.3 \, t$, which increases $s$ to 0.28, 
causes the computation to match experiment. The agreement at low-energy 
scales in either case is sufficient to demonstrate that the spin wave 
velocity is correctly computed.

\begin{figure}
\includegraphics[scale=0.4]{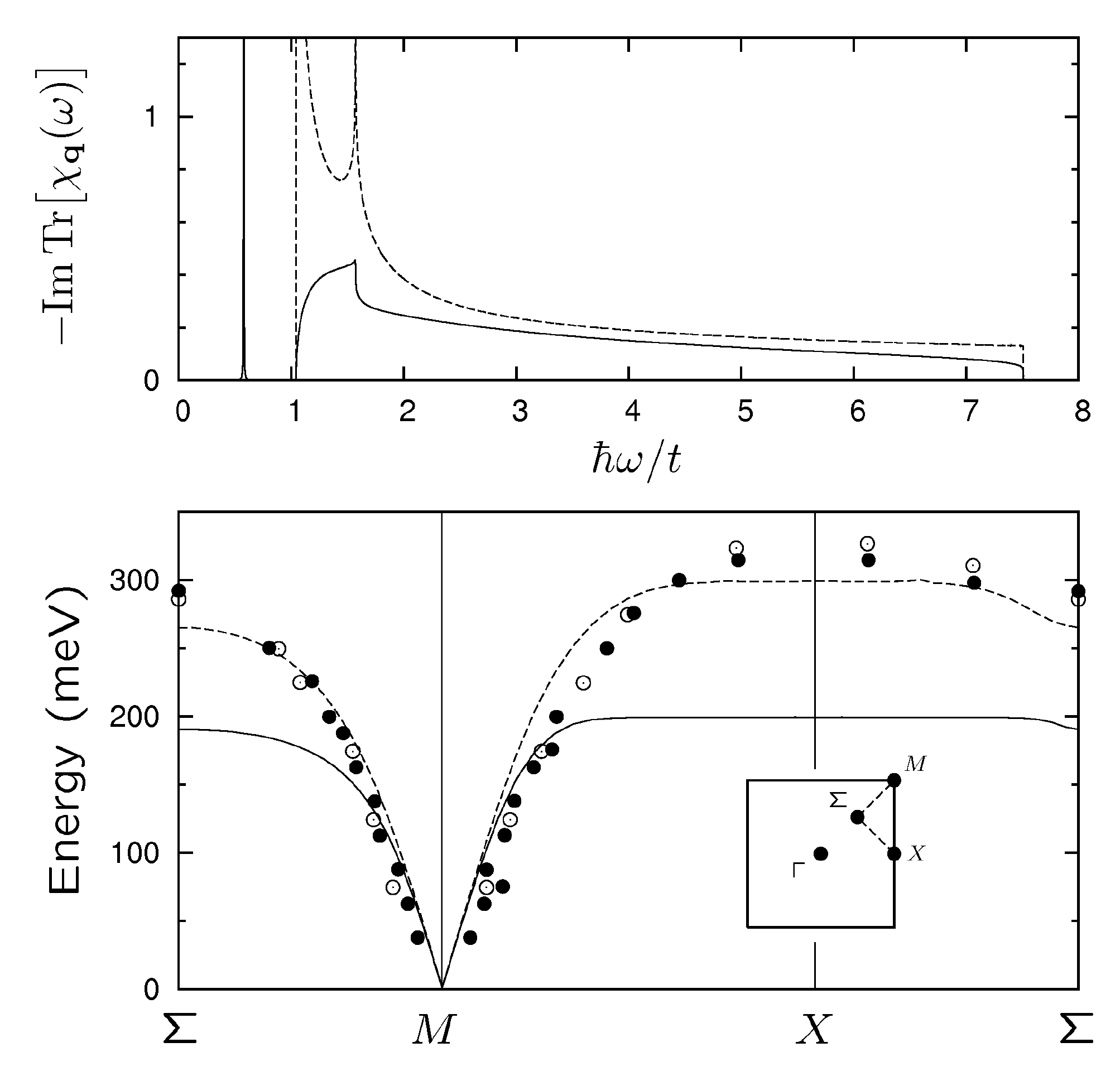}
\caption{Top: Imaginary part of the ladder sum (RPA) spin susceptibility 
defined by Eq. (\ref{rpa}) for a momentum transfer of $(0.9 \pi, 0.9 
\pi)$. The spin wave is the $\delta$-function pole at $\hbar \omega = 0.58 
\, t$ bound down below the particle-hole continuum. The dashed curve is 
the bare susceptibility defined by Eq. (\ref{chi0}). The computation 
assumes the parameters in Fig. \ref{f1}: $U = 0.76 \, t$, $J = 0.75 \, 
t$, $V_c =0.87 \, t$, $t' = 0.1 \, t$, $V_n = V_t = 0$, and $t = 0.19$ 
eV. Bottom: Comparison of magnon dispersion relation computed using 
Eqs. (\ref{rpa}) and (\ref{chi0}) with that measured by inelastic 
neutron scattering by Coldea {\it al.} for La$_2$CuO$_4$ \cite{coldea}. 
The spin wave velocity is $\hbar v_s = 1.0$ eV \AA. The open and solid 
symbols correspond to measurements taken at $T = 10$ K and $T = 
295$ K. The dashed curve is a repeat of the calculation with $U = 
1.3 \, t$ and all other parameters the same. This raises the magnetic 
moment to $s = 0.28$. The inset shows the location of the symmetry 
points in the Brillouin zone.}
\label{f12} 
\end{figure}

\subsection*{Chemical Potential Jump}

The chemical potential jump upon going from $p$ type to $n$ type is small 
\cite{harima,ikeda}. Exactly how small is not known exactly, for the 
experiments capable of measuring $\Delta \mu$ are difficult and plagued 
with the same interface chemistry sensitivity one finds in Schottky 
barrier heights. The calculation finds this jump to be twice the SDW gap 
at half filling, or 0.2 eV.  Harima {\it et al.} originally reported a 
value of between 0.15 and 0.40 eV based core level shifts measured by 
x-ray photoemission \cite{harima}. The core lines they reported were 
much broader than the shifts in question, and the core levels of 
different chemical species in the sample also shifted by different 
amounts, sometimes even in different directions.  More recent 
experiments by Ikeda {\it et al.} on samples of 
Y$_{0.38}$La$_{0.62}$Ba$_{0.74}$La$_{0.26}$Cu$_3$O$_y$, which can be 
doped chemically in both directions, showed a larger jump of 0.8 eV 
\cite{ikeda}. But both reported shifts are much smaller than the optical 
gap.

\section{Summary and Conclusion}

The theory discussed in this paper differs substantially from most 
previous theoretical work on the cuprates in being based primarily on 
fundamental quantum mechanics, not empiricism.  The underlying objective 
in constructing it is not to to make a consistent mathematical 
phenomenology but to identify basic principles that can enable one to 
distinguish correct experiments from incorrect ones. This is especially 
important in a field plagued with reproducibility and materials 
problems.  While a handful of parameters discussed in this paper are fit 
to experiment, the equations themselves are not. A system evolved out of 
a fictitious metallic parent {\it must} have a low-energy excitation 
spectrum described by the Hamiltonian ${\cal H}_0 + \Delta {\cal H}$, 
defined by Eqs. (\ref{h0}) and (\ref{dh}), regardless of details.  All 
important behaviors of the cuprates---antiferromagnetism, half-filling 
insulation, pseudogap and $d$-wave superconductivity---therefore {\it 
must} be contained in it. This is a different statement from suggesting 
yet another model for the cuprates.

A specific, and important, thing ruled out by ${\cal H}_0 + \Delta {\cal 
H}$ is Mott insulation.  A system evolved adiabatically out of a 
metallic parent could very possibly undergo a quantum phase transition 
to a new state of matter that insulates, but one is obligated to write 
down equations for what this state is in terms of the low-lying 
excitations of the metal, ordinary electrons and holes, and make the 
case that the state is stable.  A good example of such discipline is the 
Bardeen-Cooper-Schrieffer theory of superconductivity, which is a 
logical construct built on top of a metallic parent that is, in fact, 
fictional because it is absolutely unstable to superconductivity 
\cite{schrieffer}. Thus, if the Mott insulator existed it would be 
possible to derive it starting from a Hamiltonian of the form of ${\cal 
H}_0 + \Delta {\cal H}$ with a specific choice of parameters.  But this 
cannot be done. The Mott insulator therefore does not exist, and it 
cannot be used as a parent vacuum on which to build a theory of 
superconductivity.  The Varma loop current insulator must be rejected 
for the same reason.

The Hamiltonian ${\cal H}_0 + \Delta {\cal H}$ requires the pseudogap to 
be bond current antiferromagnetism, or $d$-density wave (DDW).  The 
reason is that DDW is fundamentally a crystal of $d$-wave Cooper pairs.  
It is automatically stabilized by any Hamiltonian that stabilizes 
$d$-wave superconductivity.  Moreover, there is no latitude in the 
parameter space of ${\cal H}_0 + \Delta {\cal H}$ for stabilizing other 
states, particularly ones with the $d_{x^2 - y^2}$ quasiparticle 
symmetry characteristic of the $d$-wave superconductor.  DDW order {\it 
must} exist in the pseudogap regime of the cuprates, notwithstanding the 
problematic neutron searches for its signature magnetic Bragg peaks. The 
nuclear quadrupole resonance measurements reporting an absence of 
orbital current of any kind are therefore either misinterpreted or in 
error \cite{strassle}.

The number of free parameters in Eqs. (\ref{h0}) and (\ref{dh}) is 
actually three.  The band parameters $t$ and $t'$ are fixed by band 
structure calculations and photoemission measurements on a wide variety 
of cuprates. The latter also require $V_t = 0$.  Stabilizing the 
pseudogap near half filling requires $V_n = 0$.  The three remaining 
parameters are fit to (1) the half filling antiferromagnetic moment, 
(2) the maximum pseudogap magnitude, and (3) the maximum superconducting 
$T_c$.

The parameter fit that results points unequivocally to the bonding O 
atom in the Cu-O plane as the pairing agent of high-temperature 
superconductivity. Superexchange mediated by this atom causes all three 
kinds of order: spin antiferromagnetism (SDW), $d$-density wave (DDW), 
and $d$-wave superconductivity (DWS).  Achieving proper balance among 
them requires that the atom also mediate repulsive Cooper pair 
tunneling, but this is fundamentally the same effect as 
antiferromagnetic spin exchange.

With the parameters fixed to values of Fig. \ref{f1}, solution of the 
equations with conventional Hartree-Fock methods accounts quantitatively 
for the following aspects of cuprate superconductors:

\begin{enumerate}

\item A Fermi surface congruent with the underlying density functional
      band structure.

\item Spin antiferromagnetism (SDW) at half filling with a moment of about 
      $s = 0.25$, or half the classical Ising value.

\item A spin wave velocity at half filling of $\hbar v_s \simeq 1.0$ eV 
      \AA.

\item Insulating behavior of the SDW for any value of doping.

\item Destruction of the SDW at 5\% $p$-type doping.

\item Supplanting of the SDW at this doping by a pseudogap phase (DDW)
      characterized by a $d$-wave node.

\item The simultaneous reconstruction of the Fermi surface into pockets
      compatible with observed quantum oscillations.

\item A pseudogap magnitude of approximately 110 meV that decreases
      continuously to zero at 16\% $p$-type doping.

\item An underlying quantum phase transition at 16\% doping involving
      Fermi surface reconnection and thus massive carrier scattering.

\item The coexistence of the pseudogap (DDW) with $d$-wave 
      superconductivity (DWS).

\item A $d$-wave superconducting gap maximizing at 20 meV at 16\% doping 
      and declining away from this optimal value.

\item A minimum in-plane London penetration depth of $\lambda_0
      = 0.1 \, \mu$m at optimal doping.

\item Strong suppression of the DWS superfluid density at $p$-type dopings 
      less than optimal.

\item A superconducting transition temperature $T_c$ maximizing at 93 
      K and equal at all important dopings to 1/2.2 times the maximum 
      $d$-wave gap.

\item The absence of a pseudogap for $n$-type doping.

\item A direct transition from SDW to DDW at 15\% $n$-type doping.

\item A chemical potential jump when going from $p$-type to $n$-type
      doping of approximately 0.2 eV.

\end{enumerate}

\noindent
The last of these items is sensitive to the half-filling magnetic moment 
and perhaps indicates an SDW gap that is too small.  It is, however, 
consistent with experimental reports that the jump is much smaller than 
the optical gap.

The quality of these results, the extreme simplicity of the equations 
that produced them, and the reasonableness of the parameters required all 
argue strongly in favor of the theory's physical correctness.

\section*{Acknowledgments}

I wish to thank S. Chakravarty for numerous insightful discussions about 
the cuprate problem over the years and for his determination in pursuing 
the empirical case for DDW. I also wish to thank S. Kivelson, T. 
Geballe, S.-C. Zhang, T. Devereaux and R. Martin for helpful 
discussions.  Special thanks go to S. Raghu for calling the need for 
numerical work on this problem to my attention.
  
\appendix

\section{DDW/DWS Degeneracy}

\label{phtrans}

The equivalence of Eqs. (\ref{scci}) and (\ref{scxi}) at half filling 
and $t' = 0$ results from a special symmetry familiar from studies 
of the Hubbard model \cite{nagaoka}. The Hamiltonian ${\cal H}_0 + 
\Delta {\cal H}$ defined by Eqs. (\ref{h0}) and (\ref{dh}) transforms 
simply under the unitary operator

\begin{equation}
{\cal U} = \prod_j^N \biggl[ c_{j \downarrow}^\dagger
- (-1)^j c_{j \downarrow} \biggr]
\end{equation}

\noindent
the action of which on the primary field operators is

\begin{equation}
{\cal U} c_{j \uparrow} {\cal U}^\dagger
= c_{j \uparrow}
\; \; \; \; \; 
{\cal U} c_{j \downarrow} {\cal U}^\dagger
= (-1)^j \; c_{j \downarrow}^\dagger
\end{equation}

\subsection*{Hubbard Parameter Transformations}

\noindent
The kinetic energy transforms to itself

\begin{equation}
{\cal U} \, (c_{j \uparrow}^\dagger c_{k \uparrow}
+ c_{j \downarrow}^\dagger c_{k \downarrow} ) \, {\cal U}^\dagger
= (c_{j \uparrow}^\dagger c_{k \uparrow}
+ c_{j \downarrow}^\dagger c_{k \downarrow})
\end{equation}

\noindent
The on-site repulsion negates

\begin{equation}
{\cal U} \biggl[ 
c_{j \uparrow}^\dagger c_{j \downarrow}^\dagger
c_{j \downarrow} c_{j \uparrow} \biggr] {\cal U}^\dagger
= - c_{j \uparrow}^\dagger c_{j \downarrow}^\dagger
c_{j \downarrow} c_{j \uparrow} +
c_{j \uparrow}^\dagger c_{j \uparrow}
\label{utrans}
\end{equation}

\noindent
The leftover $\sum_j c_{j \downarrow}^\dagger c_{j \downarrow}$ is a 
constant of the motion and may be dropped, which restores 
spin-rotational invariance.

\subsection*{Spin Exchange Transformation}

\noindent
The spin exchange term becomes

\begin{displaymath}
\sum_{\sigma \sigma'} {\cal U}  \biggl[
c_{j \sigma}^\dagger c_{k \sigma'}^\dagger
c_{k \sigma} c_{j \sigma'} 
- \frac{1}{2} c_{j \sigma}^\dagger c_{k \sigma'}^\dagger
c_{k \sigma'} c_{j \sigma} \biggr] {\cal U}^\dagger
\end{displaymath}

\begin{displaymath}
= \frac{1}{2}
\sum_{\sigma \sigma'} \;
c_{j \sigma}^\dagger c_{k \sigma'}^\dagger
c_{k \sigma'} c_{j \sigma}
- \biggl[
c_{j \uparrow}^\dagger c_{j \downarrow}^\dagger
c_{k \downarrow} c_{k \uparrow}
\end{displaymath}

\begin{equation}
+ c_{k \uparrow}^\dagger c_{k \downarrow}^\dagger
c_{j \downarrow} c_{j \uparrow} \biggr]	
+ \frac{1}{2} \biggl[ 1 - \sum_\sigma (
c_{j \sigma}^\dagger c_{j \sigma}
+ c_{k \sigma}^\dagger c_{k \sigma} ) \biggr]
\label{jtrans}
\end{equation}

\noindent
The last term in this expression is again a constant of the motion that 
may be dropped.  Thus, in effect, the parameters $J$ and $V_n - 2 V_c$
turn into each other under the action of ${\cal U}$.

\subsection*{Remaining Transformations}

\noindent
The action of ${\cal U}$ on $V_n$ or $V_c$ alone does not generate a 
spin invariant interaction. We have specifically

\begin{displaymath}
\sum_{\sigma \sigma'} {\cal U} \;  
c_{j \sigma}^\dagger c_{k \sigma'}^\dagger
c_{k \sigma'} c_{j \sigma} \;
{\cal U}^\dagger 
\end{displaymath}

\begin{equation}
= \sum_{\sigma \sigma'} ( c_{j \uparrow}^\dagger c_{j\uparrow} -
c_{j \downarrow}^\dagger c_{j\downarrow} + 1)
(c_{k \uparrow}^\dagger c_{k\uparrow} -
c_{k \downarrow}^\dagger c_{k\downarrow} + 1)
\end{equation}

\noindent
Thus the combination $V_n - 2 V_c$ is special.

\subsection*{Order Parameter Transformation}

\noindent
The DDW order parameter transforms under the action of ${\cal U}$ to

\begin{displaymath}
{\cal U} \biggl[ \frac{1}{2i} \sum_\sigma (
c_{j \sigma}^\dagger c_{k \sigma} -
c_{k \sigma}^\dagger c_{j \sigma} ) \biggr] {\cal U}^\dagger
\end{displaymath}

\begin{equation}
= \frac{1}{2i} \biggl[ 
(c_{j \uparrow}^\dagger c_{k \uparrow} -
c_{j \downarrow}^\dagger c_{k \downarrow} ) -
(c_{k \uparrow}^\dagger c_{j \uparrow} -
c_{k \downarrow}^\dagger c_{j \downarrow} ) \biggr]
\end{equation}

\begin{displaymath}
= \frac{1}{2 i} ( \,
< \! \Phi_j^\dagger | \sigma_z | \Phi_k \! > -
< \! \Phi_k^\dagger | \sigma_z | \Phi_j \! > \, )
\end{displaymath}

\noindent
where

\begin{displaymath}
| \Phi_j \! > = 
\left[ \begin{array}{c}
c_{j \uparrow} \\ c_{j \downarrow} \\ \end{array} \right]
\; \; \; \; \; \;
\sigma_x = 
\left[ \begin{array}{rr} 0 & 1 \\ 1 & 0 \\ \end{array} \right]
\end{displaymath}

\begin{equation}
\sigma_y = 
\left[ \begin{array}{rr} 0 & -i \\ i & 0 \\ \end{array} \right]
\; \; \; \; \; \;
\sigma_z = 
\left[ \begin{array}{rr} 1 & 0 \\ 0 & -1 \\ \end{array} \right]
\end{equation}

\noindent
This is a bond spin current polarized in the z-direction.  Thus,
if ${\cal H}_0 + \Delta {\cal H}$ stabilizes DDW then ${\cal U} \,
({\cal H}_0 + \Delta {\cal H}) \, {\cal U}^\dagger$ stabilizes
bond spin current.  But it must also stabilize spin currents in the
x-direction and y-direction because it is spin-rotationally
invariant.  Back-transforming the latter we find

\begin{displaymath}
{\cal U}^\dagger \biggl[ 
\frac{1}{2 i} ( \,
< \! \Phi_j^\dagger | \sigma_x | \Phi_k \! > -
< \! \Phi_k^\dagger | \sigma_x | \Phi_j \! > \, ) \biggr]
{\cal U}
\end{displaymath}

\begin{equation}
= - \frac{(-1)^j}{2 i} \biggl[
(c_{k \uparrow}^\dagger c_{j \downarrow}^\dagger
+ c_{j \uparrow}^\dagger c_{k \downarrow}^\dagger) -
(c_{k \downarrow} c_{j \uparrow}
+ c_{j \downarrow} c_{k \uparrow}) \biggr]
\end{equation}

\begin{displaymath}
{\cal U}^\dagger \biggl[ 
\frac{1}{2 i} ( \,
< \! \Phi_j^\dagger | \sigma_y | \Phi_k \! > -
< \! \Phi_k^\dagger | \sigma_y | \Phi_j \! > \, ) \biggr]
{\cal U}
\end{displaymath}

\begin{equation}
= (-1)^j \biggl[
(c_{k \uparrow}^\dagger c_{j \downarrow}^\dagger
+ c_{j \uparrow}^\dagger c_{k \downarrow}^\dagger) +
(c_{k \downarrow} c_{j \uparrow}
+ c_{j \downarrow} c_{k \uparrow}) \biggr]
\end{equation}

\noindent
These represent the real and imaginary parts of the $d$-wave 
superconductor order parameter.  Thus stabilizing DDW with $U$ and 
$J$ solely at half filling necessarily stabilizes DWS also.

\section{Spiral States}

\label{spiraltrap}

The nature of SDW spirals is most easily illustrated by transforming to 
a basis in which each unit cell is the same.  For simplicity we set all 
the parameters in ${\cal H} = {\cal H}_0 + \Delta{\cal H}$ to zero 
except for $t$, $U$ and $J$. The unitary transformation

\begin{equation}
{\cal U}_{\bf Q} = \prod_j^N \exp \biggl\{ 
\frac{i}{2} \,
({\bf Q} \cdot {\bf r}_j) \, 
\biggl[ c_{j \uparrow}^\dagger c_{j \downarrow} +
c_{j \downarrow}^\dagger c_{j \uparrow} \biggr] \biggr\}
\end{equation}

\noindent
spirals the spin direction in the $yz$ plane as one advances in the 
direction of ${\bf Q}$.  Its actions on the elementary field operators 
are

\begin{equation}
{\cal U}_{\bf Q} c_{j \uparrow}^\dagger {\cal U}_{\bf Q}^\dagger
= \cos( \frac{{\bf Q} \cdot {\bf r}_j}{2}) \, c_{j \uparrow}^\dagger
+ i \, \sin( \frac{{\bf Q} \cdot {\bf r}_j}{2}) \, 
c_{j \downarrow}^\dagger
\end{equation}

\begin{equation}
{\cal U}_{\bf Q} c_{j \downarrow}^\dagger {\cal U}_{\bf Q}^\dagger
= \cos( \frac{{\bf Q} \cdot {\bf r}_j}{2}) \, 
c_{j \downarrow}^\dagger
+ i \, \sin( \frac{{\bf Q} \cdot {\bf r}_j}{2}) \, 
c_{j \uparrow}^\dagger
\end{equation}

\noindent
The ansatz of a single-Slater-determinant ground state periodic in the 
twisted unit cell gives the Hartree-Fock Hamiltonian

\begin{equation}
{\cal H}_{\bf q}^{HF} 
= \epsilon_{\bf q}^{(0)} \,
\left[ \begin{array}{rr} 1 & 0 \\ 0 & 1 \end{array} \right] 
+ \epsilon_{\bf q} \,
\left[ \begin{array}{rr} 0 & 1 \\ 1 & 0 \end{array} \right] 
+ \Delta \, 
\left[ \begin{array}{rr} 1 & 0 \\ 0 & -1 \end{array} \right] 
\end{equation}

\noindent
where

\begin{displaymath}
\epsilon_q =
2 t \biggl[ \sin(\frac{Q_x}{2}) \sin(q_x)
+ \sin(\frac{Q_y}{2}) \sin(q_y) \biggr]
\end{displaymath}

\begin{displaymath}
\epsilon_{\bf q}^{(0)} =
- 2 t \biggl[ \cos(\frac{Q_x}{2}) \cos(q_x)
+ \cos(\frac{Q_y}{2}) \cos(q_y) \biggr]
\end{displaymath}

\noindent
The corresponding quasiparticle dispersion relation is 
$\epsilon_{\bf q}^{(0)} \pm E_{\bf q}$, where

\begin{equation}
E_{\bf q} = \sqrt{ \epsilon_{\bf q}^2 + \Delta^2}
\end{equation}

\noindent
When $Q_x$ and $Q_y$ both equal $\pi$, this reduces to the doubled unit 
cell case worked out in Section \ref{hartreefock}.  We are interested in 
$Q_x$ and $Q_y$ both near $\pi$, but not equal to it.  The SDW band gap 
is then preserved, and the self-consistency equation for $\Delta$ 
becomes

\begin{equation}
\Delta = ( U - J [ \cos(Q_x) + \cos(Q_y)]) s
\end{equation}

\noindent
where

\begin{equation}
s = \frac{1}{8 \pi^2}
\int_{-\pi}^\pi \int_{-\pi}^\pi
\frac{\Delta}{E_{\bf q}} \, dq_x dq_y
\end{equation}

\begin{equation}
\chi_R = \frac{1}{32 \pi^2 t}
\int_{-\pi}^\pi \int_{-\pi}^\pi
\frac{\epsilon_{\bf q}^2}{E_{\bf q}}
\end{equation}

\noindent
The total expected energy is

\begin{displaymath}
\frac{< \! {\cal H} \! >}{N}
= - 8 \chi_R t + (\frac{1}{4} - s^2 ) \, U 
\end{displaymath}

\begin{equation}
+ [ \cos(Q_x) + \cos(Q_y) ] s^2 \, J
\end{equation}

\noindent
The effect of twisting on the corresponding density of states

\begin{equation}
{\cal D}(E) = \frac{1}{8 \pi^2} \sum_\pm
\int_{-\pi}^\pi \int_{-\pi}^\pi
\delta ( E - \epsilon_{\bf q}^{(0)} \pm E_{\bf q}) \,
dq_x dq_y
\label{spiraldos}
\end{equation}

\noindent
is shown in Fig. \ref{f13}.  At a very small cost in variational energy, 
a spiral twist narrows the band gap, thus lowering the energy of added 
carriers.  This implies that twisted antiferromagnetic domain walls trap 
carriers and cause insulation.

\begin{figure}
\includegraphics[scale=0.4]{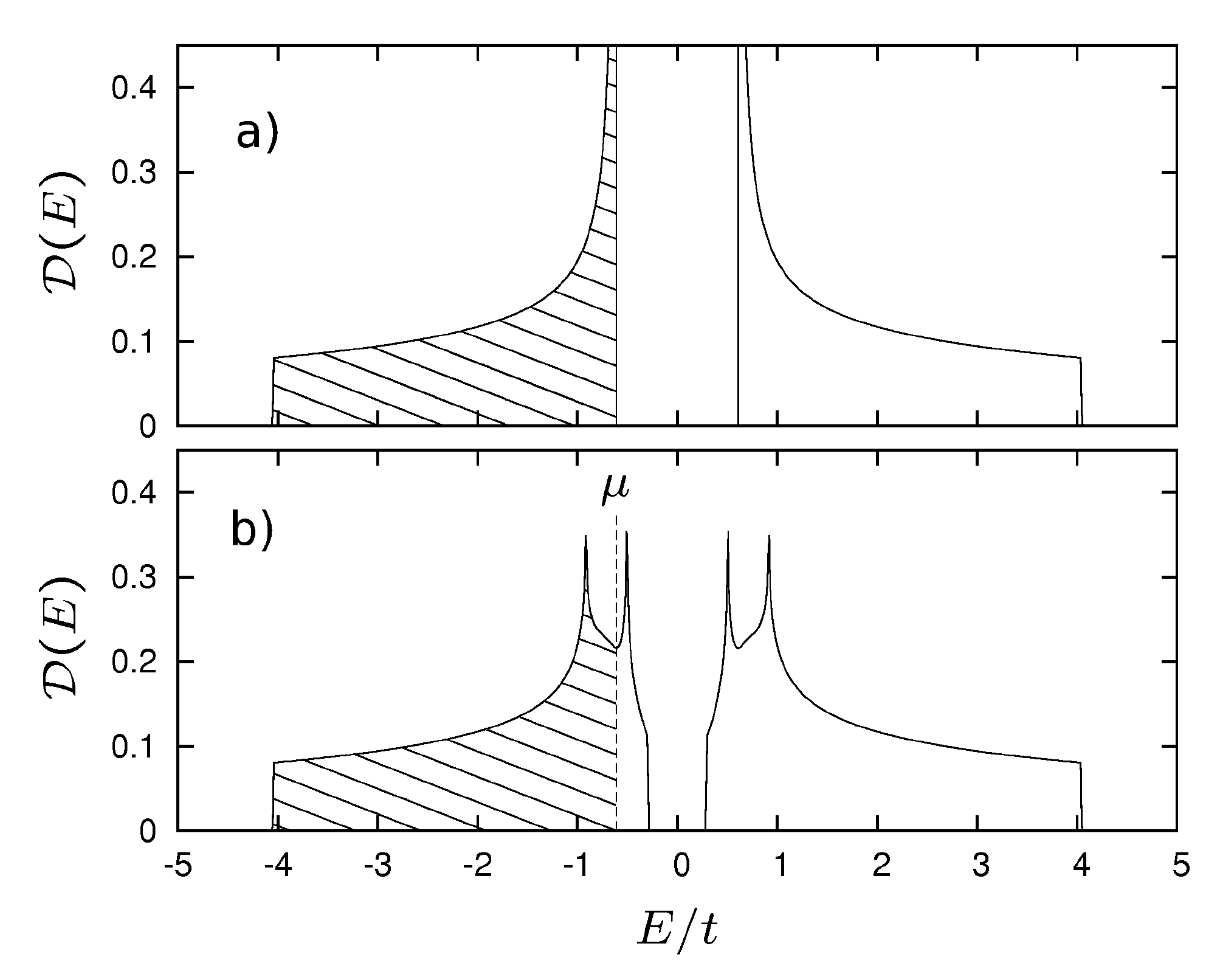}
\caption{Density of states for spiral state defined by Eq. 
(\ref{spiraldos}) for the case of $J = 0.7 \, t$ and $U = 1.1 \, t$. (a) 
Untwisted state: $(Q_x, Q_y) = (\pi, \pi)$, $< \! {\cal H} \! >/N = 
-1.391 \, t$.  (b) Twisted state: $(Q_x , Q_y) = (0.9 \pi, \pi)$, $< \! 
{\cal H} \! >/N = 1.380 \, t$.  As a practical matter, the spiral state 
represents a domain wall. This traps carriers added to the system the 
way defects in a semiconductors do, thus causing the system to 
insulate.}
\label{f13}
\end{figure}

\section{Optical Sum Rule}

\label{fsumrule}

The $f$-sum rule for lattice Hamiltonians is widely discussed in the 
literature, but let us reprise it briefly \cite{monien}. The buildup of 
electron density in response to applied potentials is described to 
linear order by density-density response function

\begin{equation}
< \! \rho_{-{\bf q}} \, \rho_{\bf q} \! >
= \sum_x | < \! x | \rho_{\bf q} | 0 \! > |^2 
\frac{2 E_x}{(\hbar \omega + i \eta)^2 - E_x^2}
\end{equation}

\noindent
where $| x \! >$ and $| 0 \! >$ are exact excited and ground states
of the Hamiltonian ${\cal H}$ with energy eigenvalues $E_x$
and $E_0 = 0$, $\eta$ is an infinitesimal, and

\begin{equation}
\rho_q = \sum_j^N \sum_\sigma \exp(i {\bf q} \cdot {\bf r}_j ) \;
c_{j \sigma}^\dagger c_{j \sigma}
\end{equation}

\noindent
The sum rule is

\begin{displaymath}
\int_0^\infty {\rm Im} 
< \! \rho_{-{\bf q}} \, \rho_{\bf q} \! > 
\omega d\omega 
= -  \frac{\pi}{2 \hbar^2}
< \! 0 |  [ \rho_{-{\bf q}} , [{\cal H} , \rho_{\bf q}] ] | 0 \! > 
\end{displaymath}

\begin{displaymath}
= - \frac{\pi}{2 \hbar^2} \biggl\{
t \sum_{< \! j k \! >}^{2N} | e^{i {\bf q} \cdot {\bf r}_j}
- e^{i {\bf q} \cdot {\bf r}_k} |^2 \sum_\sigma
(c_{j \sigma}^\dagger c_{k \sigma} + c_{k \sigma}^\dagger 
c_{j \sigma} )
\end{displaymath}

\begin{equation}
- t' \sum_{< \! j \ell \! >}^{2N} | e^{i {\bf q} \cdot {\bf r}_j}
- e^{i {\bf q} \cdot {\bf r}_\ell} |^2 \sum_\sigma
(c_{j \sigma}^\dagger c_{\ell \sigma} + c_{\ell \sigma}^\dagger 
c_{j \sigma} ) \biggr\}
\end{equation}

\noindent
We thus have

\begin{equation}
\lim_{{\bf q} \rightarrow 0} \frac{1}{N q^2 b^2}
< \! 0 |  [ \rho_{-{\bf q}} , [{\cal H} , \rho_{\bf q}] ] | 0 \! > 
= 4 \chi_R t - 8 \chi_R' t'
\end{equation}

\noindent
Contact with experiment is made through the in-plane longitudinal
conductivity (in gaussian units)

\begin{equation}
\sigma(\omega) = i \omega \lim_{{\bf q} \rightarrow 0}
\frac{e^2}{N a b^2 q^2} < \! \rho_{-{\bf q}} \rho_{\bf q} \! >
\end{equation}

\end{document}